\documentclass[12pt,letterpaper,journal,onecolumn]{IEEEtran}
\usepackage{amsmath,amssymb,amsthm}
\newtheorem{definition}{Definition}
\usepackage{enumitem}
\usepackage{algorithmic}
\usepackage{array}
\usepackage[caption=false,font=normalsize,labelfont=sf,textfont=sf]{subfig}
\usepackage{textcomp}
\usepackage{stfloats}
\usepackage{url}
\usepackage{verbatim}
\usepackage{graphicx}
\def\BibTeX{{\rm B\kern-.05em{\sc i\kern-.025em b}\kern-.08em
    T\kern-.1667em\lower.7ex\hbox{E}\kern-.125emX}}

\usepackage[utf8]{inputenc}
\usepackage[english]{babel}
\usepackage{xcolor}
\usepackage{cite}
\usepackage[T1]{fontenc} 
\usepackage{tabularx}
\usepackage{ragged2e}
\usepackage{multirow}
\usepackage{bm}
\usepackage{tikz}
\usetikzlibrary{arrows.meta, positioning, calc}
\usepackage{mathrsfs}
\usepackage{psfrag}
\usepackage{float}
\usepackage{mathtools}
\usepackage{amssymb}
\usepackage{algorithm}
\usepackage[normalem]{ulem}
\usepackage{soul}
\usepackage{comment}
\usepackage{booktabs}
\usepackage{lipsum}
\usepackage{mathtools}
\usepackage[acronym]{glossaries}
\usepackage{balance}
\usepackage{siunitx}
\usepackage{setspace}
\newcommand{\TransmittedData}{d^j_{n,t}}
\newcommand{\BatteryStateDynamics}{b^j_t}
\newcommand{\QueueStateDynamics}{q^j_{n,t}}

\newcommand{\HarvestedPower}{p^\prime_{n,t}}

\newcommand{\TransmissionPower}{p^j_{n,t}}
\newcommand{\PrimaryMaxPower}{P_0}
\newcommand{\SecondaryMaxPower}{P_n}
\newcommand{\PrimaryTransmissionPower}{p^j_{0,t}}

\newcommand{\CommunicationTime}{\alpha^j_{n,t}}
\newcommand{\PrimaryCommunicationTime}{\alpha^j_{0,t}}

\newcommand{\SecondaryCommunicationTime}{\alpha^j_{n,t}}
\newcommand{\EHTime}{{\alpha^\prime}^j_{n,t}}
\newcommand{\tx}{\text{TX}}
\newcommand{\rx}{\text{RX}}
\newcommand{\txPrimary}{\tx_0}
\newcommand{\txSecondary}{\tx_n}
\newcommand{\rxPrimary}{\rx_0}
\newcommand{\rxSecondary}{\rx_n}
\newcommand{\primary}{(\txPrimary,\rxPrimary)}
\newcommand{\secondary}{(\txSecondary,\rxSecondary)}
\newcommand{\txI}{\tx_n}
\newcommand{\rxI}{\rx_n}
\newcommand{\systemI}{(\txI,\rxI)}
\newcommand*{\starnr}{\stepcounter{equation}\tag{\theequation}}


\begin{document}

% acronym

% \newacronym[plural=cMs,firstplural=centiMorgans (cMs)]{cM}{cM}{centiMorgan}
\newacronym{EH}{EH}{energy harvesting} 
\newacronym{WSN}{WSN}{wireless sensor network} 
\newacronym{WSNs}{WSNs}{wireless sensor networks}
\newacronym{MDP}{MDP}{Markov decision process} 
\newacronym{L2RL}{L2RL}{lifelong reinforcement learning} 
\newacronym{CD-L2RL}{CD-L2RL}{cross-domain lifelong reinforcement learning} 
\newacronym{RLman}{RLman}{energy management algorithm based on reinforcement learning}
\newacronym{SARSA}{SARSA}{state-action-reward-state-action}
\newacronym{RF}{RF}{radio-frequency}
\newacronym{RL}{RL}{reinforcement learning} 
\newacronym{UAV}{UAV}{unmanned aerial vehicle} 
\newacronym{AoI}{AoI}{age of information} 
\newacronym{IoT}{IoT}{internet of things} 
\newacronym{SDMT-L2RL}{SDMT-L2RL}{single-domain multi-tasks L2RL} 
\newacronym{SWIPT}{SWIPT}{simultaneous wireless information and power transfer} 
\newacronym{PPP}{PPP}{poisson point process} 
\newacronym{PEARL}{PEARL}{protocol for energy-efficient adaptive scheduling using reinforcement learning}
\newacronym{PG}{PG}{policy gradient} 
\newacronym{WET}{WET}{wireless energy transfer} 
\newacronym{KKT}{KKT}{Karush-Kuhn-Tucker} 
\newacronym{POMDP}{POMDP}{partially observable Markov decision process} 
\newacronym{SNR}{SNR}{signal-to-noise ratio} 
\newacronym{SER}{SER}{symbol error rate} 
\newacronym{AWGN}{AWGN}{additive white gaussian noise} 
\newacronym{MT-L2RL}{MT-L2RL}{multi-tasks L2RL}  
\newacronym{NN}{NN}{neural network}  
\newacronym{TX}{TX}{transmitter}
\newacronym{RX}{RX}{receiver}
\newacronym{ML}{ML}{machine learning} 
\newacronym{PDF}{PDF}{probability density function}
\newacronym{KB}{KB}{knowledge base}
\newacronym{QoS}{QoS}{quality-of-service}
\newacronym{URLLC}{URLLC}{ultra-reliable and low-latency communication}
\newacronym{NLOS}{NLOS}{non-line-of-sight}
\newacronym{CDF}{CDF}{cumulative distribution function}
\newacronym{TD}{TD}{temporal-difference}
\newacronym{LASSO}{LASSO}{least-absolute-shrinkage-and-selection-operator}

\title{Cross-Domain Lifelong Reinforcement Learning for Wireless Sensor Networks}
\author{
\IEEEauthorblockN{
Hossein~Mohammadi~Firouzjaei\IEEEauthorrefmark{1}, %~\IEEEmembership{Member,~IEEE,}
Rafaela~Scaciota\IEEEauthorrefmark{1}\IEEEauthorrefmark{2}, %~\IEEEmembership{Member,~IEEE,} 
Sumudu~Samarakoon\IEEEauthorrefmark{1}\IEEEauthorrefmark{2}, %~\IEEEmembership{Member,~IEEE}
and Beatriz~Lorenzo\IEEEauthorrefmark{3} %~\IEEEmembership{Member,~IEEE}
} \\
\IEEEauthorblockA{
	\small%
	\IEEEauthorrefmark{1}%
	Centre for Wireless Communication, University of Oulu, Finland \\
    \IEEEauthorrefmark{2}%
    Infotech Oulu, University of Oulu, Finland \\
    \IEEEauthorrefmark{3}%
    Department of Electrical and Computer Engineering, University of Massachusetts Amherst, Amherst, MA 01003, USA \\
	Email: \{hossein.mohammadifirouzjaei, rafaela.scaciotatimoesdasilva, sumudu.samarakoon\}@oulu.fi, blorenzo@umass.edu
}
\vspace{-20pt}
\thanks{
This paper benefited from language revision assistance provided by ChatGPT, a language model developed by OpenAI. 
The authors take full responsibility for the content of the manuscript.}
}

\maketitle

\begin{abstract}
\Glspl{WSN} with \gls{EH} are expected to play a vital role in intelligent $6$G systems, particularly in applications such as industrial sensing and control, where continuous operations and sustainable energy management are essential.
Given their limited energy resources, \glspl{WSN} require energy-efficient operations to maintain long-term performance.
Their deployment in real-world scenarios is challenged by highly dynamic environments, where \gls{EH} conditions, network scale, and arrival data rate changes over time.
In this work, we consider system dynamics that yield different learning tasks, where the space of decision variables remains fixed but the strategies are variable, and learning domains, where both the decision space and the strategies evolve over time.
A scenario in which a design based on the \gls{CD-L2RL} framework is required.
More specifically, we focus on an energy-efficient \gls{WSN} design under changing \gls{EH} conditions and network scale. 
Towards this, we propose a \gls{CD-L2RL} algorithm that leverages prior experience to accelerate adaptation across multiple tasks and domains.
Unlike conventional methods, such as modeling the system with \gls{MDP} or Lyapunov optimization solution, which often assume fixed or slowly changing environments, our solution exhibits rapid policy adaptation by reusing knowledge from previously encountered tasks and domains that ensure continuous operations.
We validate the effectiveness of our approach through extensive simulations under different environmental settings. 
The results show that the proposed solution enables faster adaptations up to $35\%$ compared to the standard \gls{RL} method and up to $70\%$ compared to the Lyapunov-based optimization methods. 
It also achieves a higher total amount of harvested energy, demonstrating strong potential for deployment in dynamic $6$G \glspl{WSN}.
\end{abstract}
\begin{IEEEkeywords}
Wireless sensor networks, energy harvesting, cross-domain lifelong reinforcement learning
\end{IEEEkeywords}

\section{Introduction}

\Glspl{WSN} are increasingly being deployed in several real-world applications, ranging from environmental and industrial monitoring to healthcare and smart cities~\cite{Lu.2013}. 
These systems consist of spatially distributed sensors that collect and transmit data, often operating under strict energy constraints. To reduce energy limitations, \gls{EH} techniques have been integrated into \glspl{WSN}, enabling sensors to capture ambient energy from sources such as solar, wind, or \gls{RF} signals~\cite{Yadav.2017}. 
Among these methods, \gls{RF}-based \gls{EH} is particularly promising, as it supports \gls{SWIPT}, thereby enhancing overall energy efficiency~\cite{kumar.2022}. 
With the advent of \gls{EH}-enabled $6$G and buffer-assisted communications, the complexity and scale of \glspl{WSN} are growing, requiring highly adaptive and energy-aware control mechanisms. 
Current systems often assume stable and predictable \gls{EH} and communication conditions~\cite{Qiu.2019}, which do not reflect the real-world dynamics where energy sources, channel conditions, and sensor configurations change over time.
Optimization methods and traditional learning-based approaches are the main existing paradigms to improve the performance of \gls{EH}-\gls{WSN}.

Traditional optimization methods have played a key role in designing energy-efficient control frameworks for \gls{EH}-powered \glspl{WSN}, employing techniques such as water-filling algorithms and Lyapunov optimization to balance energy consumption and queue stability~\cite{Ulukus.2015, Ozel.2011}. 
These methods offer solid theoretical guarantees but often rely on idealized assumptions, including stationary environments and full knowledge of system dynamics. 
As a result, their applicability in practical scenarios characterized by unpredictable energy arrivals, time-varying channel conditions, and evolving network topologies remains limited. 
This has motivated the exploration of learning-based approaches capable of adapting to such uncertainties~\cite{Ortiz.2017}. 
Within this context, the core problem addressed in this paper is the design of an adaptive and efficient control framework that minimizes long-term energy consumption while maintaining queue stability under non-stationary system dynamics.

Towards modeling and solving \gls{WSN} design problems, the tools from optimization theory are utilized in the early works~\cite{Ulukus.2015, Ozel.2011, Wang.2015, Zhang.2019, Cui.2012, Hu.2018, Amirnavaei.2016, Qiu.2018', qiu.2018}.
Works on optimization-based methods focused on energy-efficient transmission scheduling using water-filling algorithms~\cite{Ulukus.2015}. 
These include directional, dynamic, and iterative variants that allocate energy optimally over time to maximize throughput and minimize delay in single- and multi-user scenarios~\cite{Ozel.2011, Wang.2015}. 
These techniques extend to complex environments like \gls{SWIPT} and heterogeneous networks~\cite{Zhang.2019}. 
In these approaches, full knowledge of the system parameters is needed to compute globally optimal decisions over time and thus are inherently offline.
To overcome this, the Lyapunov optimization emerged as a powerful online alternative. 
It converts long-term stochastic control objectives into real-time decisions through drift-plus-penalty techniques~\cite{Cui.2012}, avoiding the computational cost of dynamic programming. 
The objectives are translated into decisions by minimizing a bound on the drift-plus-penalty expression at each time slot, guiding actions that balance performance and stability in real time.
Applications include \gls{EH} relay networks~\cite{Hu.2018}, power control under unknown energy arrivals~\cite{Amirnavaei.2016}, and queue-aware transmission scheduling~\cite{Qiu.2018', qiu.2018}, all of which demonstrate that the Lyapunov-based policies can ensure system stability with low complexity.

Building on the limitations of classical optimization approaches, researchers have employed \glspl{MDP} to model the stochastic and partially observable nature of \gls{EH}-\glspl{WSN}.
\glspl{MDP} offers a principled way to learn policies that adapt to real-time energy, battery, and channel conditions~\cite{Li.2015}. 
Several works have applied \glspl{MDP} to optimize outage probability~\cite{Li.2016}, minimize transmission errors via \gls{POMDP}~\cite{Yadav.2017}, and jointly consider asynchronous channel and energy dynamics~\cite{Gong.2018}. 
Further extensions target \gls{EH} relays~\cite{Li.2017}, cooperative transmission policies~\cite{Ku.2015'}, and adaptive modulation strategies~\cite{Ku.2015}. 
These \gls{MDP}-modeled problems are typically solved using \gls{RL} algorithms when system dynamics are unknown or intractable.
Although modeling problems as \glspl{MDP} enhances adaptivity compared to static optimization, they face scalability issues and often require retraining when environmental conditions change, making them less suitable for lifelong deployment in dynamic networks.

Unlike classical \gls{MDP} solvers, model-free methods eliminate the need for known dynamics, though later model-based \gls{RL} approaches reintroduced environment models to improve sample efficiency and planning.
%More recently, \gls{RL} has been introduced to overcome the model-dependence and inflexibility of \glspl{MDP}. 
\gls{RL}-based approaches such as \gls{RLman}~\cite{Aoudia.2018}, \gls{SARSA}-based energy controllers~\cite{Ortiz.2017}, and Q-learning schedulers~\cite{Atallah.2017} have enabled nodes to learn optimal control policies through trial and error. 
These methods work well without prior knowledge of system dynamics and have shown effectiveness in enhancing throughput and energy efficiency~\cite{Blasco.2013, Cao.2019, Wei.2018, Timothy.2016}. 
However, conventional \gls{RL} algorithms struggle to generalize across tasks deployed in domains with evolving state/action spaces or non-stationary system dynamics, common features in \gls{EH}-\glspl{WSN}.
This lack of transferability and adaptability leads to long retraining times and unstable performance. 

To overcome these limitations, we have introduced \emph{multi-task} lifelong \gls{RL} paradigm in our recent work~\cite{Firouzjaei.2025}, where each task represents a single learning problem under a fixed environmental configuration
\emph{Multi-task learning} assumes that all tasks come from the same domain, i.e., they share common state-action space structures, where domains are higher-level abstraction that group together multiple structurally related tasks.
It focuses on jointly learning policies for these related tasks to improve generalization. 
Typically, task-specific policies are expressed as combinations of shared basis components drawn from a global knowledge source~\cite{Ammar.2014,kumar.2012}.
In contrast, \emph{cross-domain learning} enables knowledge transfer between heterogeneous tasks belonging to different domains. 
These tasks may have different state or action space dimensions and cannot be represented in a shared input-output space directly. 
To address this, cross-domain methods learn domain-specific projections that map local task representations into a shared latent space, where transferable policy or value function components can be reused~\cite{Ammar.2015,Han.2012}. 
This projection mechanism enables multi-task learning between structurally dissimilar tasks. 
In both paradigms, it is crucial that each task retains its own reward function to reflect distinct operational objectives, especially in cross-domain settings where environmental and structural variations are more substantial.

Building upon these insights, our work proposes a \gls{CD-L2RL} framework that learns a transferable knowledge base and adapts policies efficiently to new environments through trial and error.  
This approach ensures fast convergence, stable control, and robust performance under dynamic \gls{EH}-\gls{WSN} conditions.
This work frames dynamic \gls{EH} conditions, channel scales, and data traffic setting of \gls{WSN} operations as a multi-task \gls{RL} problem, where non-constant number of energy harvesters results in introducing some learning domains that group together multiple structurally related tasks. 
This enables continual adaptation and warm-start over tasks and domains without retraining from scratch.
%These tasks are grouped into domains that share structural similarities, and the overall problem is addressed within a \gls{CD-L2RL} framework that enables efficient knowledge reuse and rapid adaptation across tasks and domains.
To the best of our knowledge, this is the first work to realize \gls{CD-L2RL} for \gls{EH}-powered \glspl{WSN}, supporting knowledge transfer across tasks and domains while ensuring policy continuity and stability over time.

The major contributions of this paper are as follows
\begin{itemize}
    \item \textbf{Non-stationary \gls{EH}-\gls{WSN} design:} We model a \gls{WSN} with \gls{EH} nodes operating under dynamic, non-stationary conditions. 
    The system's objective is to minimize energy consumption while ensuring queue stability through optimal scheduling and power control. 
    This problem is framed as a sequence of tasks.
    
    \item \textbf{\gls{CD-L2RL} formulation for \gls{EH}-\gls{WSN}:} We frame the dynamic operations of \gls{EH}-\glspl{WSN} as a \gls{CD-L2RL} problem. 
    Each task of a single-domain represents a single learning problem under a fixed number of energy harvesters, while the \gls{EH} conditions, channel scales, and arrival data rate changes over tasks.
    Cross-domain learning paradigm represents the learning problem under different number of energy harvesters.

    \item \textbf{The development of the \gls{CD-L2RL} algorithm:} Our cross-domain knowledge transfer mechanism enables each task policy to be expressed as a sparse combination of globally shared latent components projected into domain-specific spaces. 
    This structure promotes both intra-domain reuse and cross-domain generalization. 
    By continuously updating shared and domain-specific representations, the agent incrementally builds transferable knowledge for efficient lifelong learning.

    \item \textbf{Empirical validation of transfer efficiency:} We present simulation-based evidence demonstrating that the proposed framework accelerates convergence and improves generalization across tasks and domains. 
    Our results show reduced training time and enhanced queue stability and energy efficiency in dynamic \gls{EH}-\gls{WSN} environments.
\end{itemize}

The rest of the paper is structured as follows. 
Section \ref{sec:system_model} outlines the system model and problem formulation. 
The optimization-based solution is defined in section \ref{sec:the Lyapunov_benchmark}. 
Section \ref{sec:RL_benchmark} introduces the \gls{RL}-based solution. 
Our proposed solution is given in section \ref{sec:Proposed_solution}. 
Simulation results are shown in section \ref{sec:simulation_results}. 
Lastly, Section \ref{sec:conclusion_results} provides the conclusions.

\noindent
\textit{Notation:} 
Vectors are in lowercase and bold. 
Matrices are uppercase bold.
$\mathbf{I}$ refers to the identity matrix. 
$\text{vec}(\mathbf{A})$ refers to vectorization of $\mathbf{A}$. 
Sets are calligraphic $\mathcal{X}=\{1,\ldots,X\}$ but universal sets have own notations (integers $\mathbb{Z}$, real set $\mathbb{R}$).
Expectation is $\mathbb{E}[\cdot]$.

%\sps{\textbf{Notation}:}{explain your logical reasoning on your choice of notation here. I find they are quite hard to follow.}

\section{System Model \& Problem Formulation}\label{sec:system_model}

\begin{figure}[t!]
\centering
\includegraphics[width=0.9\textwidth]{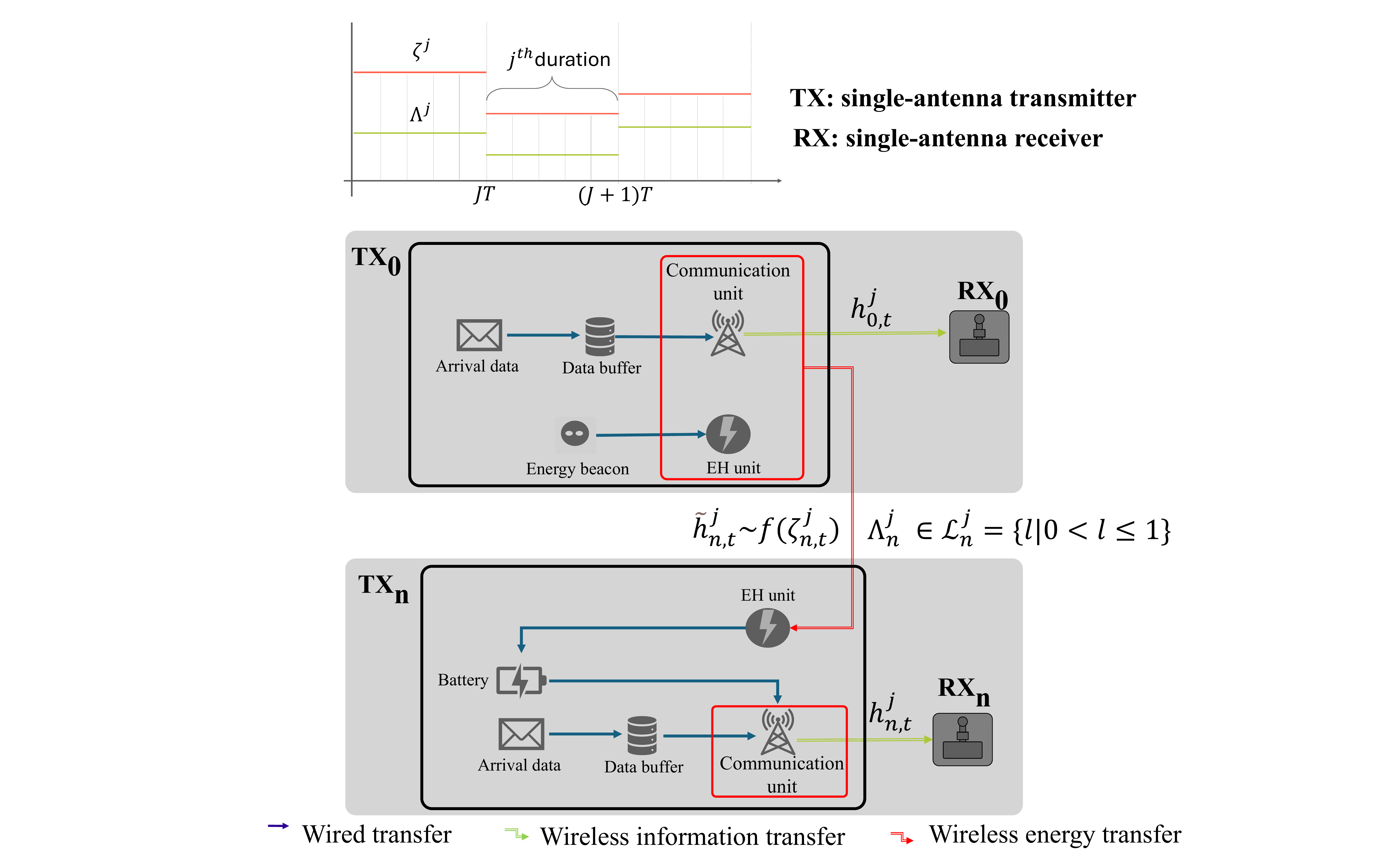}
\caption{System model comprising wireless sensors, $\primary$ and $\secondary$, in a non-stationary environment.}
\label{fig:system model}
\end{figure}

We consider a \gls{WSN} consisting of a primary sensor system and a set $\mathcal{N}$ of $N$ secondary wireless sensor systems, denoted by $\primary$ and ${\secondary}_{n \in \mathcal{N}}$, respectively.
Each sensor system is equipped with a single-antenna \gls{TX} for data transmission and a single-antenna \gls{RX} for data reception, as illustrated in Fig.~\ref{fig:system model}. 
In the primary system, $\txPrimary$ is connected to a stable energy source and transmits data to $\rxPrimary$, simultaneously serving as a power beacon for each $\txSecondary$. 
Each $\txSecondary$ is equipped with a rechargeable battery and harvests energy wirelessly from $\txPrimary$ through an \gls{EH} channel. 
Communication among each $\secondary$ is powered by the energy harvested from $\txPrimary$.
The details of the communication and wireless energy transfer models are discussed next.

We assume that the communication systems follow queuing models under channel dynamics and limited communication resources.
For any $\tx_n$ at time slot $t$, the queue state dynamics is given by~\cite{qiu.2018}
\begin{equation}\label{eqn:queue_dynamics}
    q_{n,t+1} = \max \{ q_{n,t} - d_{n,t}, 0 \} + a_{n,t},
\end{equation}
where $d_{n,t}$ is the transmitted data and $a_{n,t}$ is the arrival data, which is distributed according to a homogeneous \gls{PPP} with density $\lambda_a$ ($\si{\kilo\bit\per\second}$).
Considering that a fraction $\alpha_{n,t}$ of the time slot is used for data communication, the transmitted data is given by
\begin{equation}\label{eqn:transmitted_data}
    d_{n,t} = W \log_2 \left(1 + \frac{p_{n,t} h_{n,t}}{N_0}\right) \alpha_{n,t},
\end{equation}
where $W$ is the transmission bandwidth, $N_0$ denotes the power spectral density of \gls{AWGN}, and $p_{n,t}$ is the transmission power of $\tx_n$. 
Here, $h_{n,t}$ represents the norm of the channel gain between the $\systemI$ pair, which is modeled using a Rayleigh distribution with the scale parameter $\zeta_n > 0$. 
Considering that the remaining power at $\txPrimary$ after its data transmission is equally distributed among all the secondary systems, the instantaneous power harvested at $\txSecondary$ is given by~\cite{wang.2021}
\begin{equation}\label{eqn:harvested_power}
    \HarvestedPower = \lambda_n \left(\frac{\PrimaryMaxPower - p_{0,t}}{N}\right) h^\prime_{n,t},
\end{equation}
where $\lambda_n \in (0,1]$ is the power conversion efficiency factor and $\PrimaryMaxPower$ is the maximum available power for $\txPrimary$.
Here, $h^\prime_{n,t}$ denotes the norm of the \gls{EH} channel gain between $\txPrimary$ and $\txSecondary$, which is assumed to follow a Rayleigh fading model parameterized by $\zeta^\prime_n$.
Then, at $\txSecondary$, the harvested energy over a fraction $\alpha^\prime_{n,t}$ of the time slot is stored in its battery.
The energy stored at the battery is used for the data transmission following~\eqref{eqn:transmitted_data}.
In this view, the battery state $b_{n,t}$ dynamics of $\txSecondary$ can be modeled as
\begin{equation}
b_{n,t+1} = b_{n,t} \
- \underbrace{p_{n,t} \alpha_{n,t}}_{\text{energy used for transmission}} \
+ \underbrace{\HarvestedPower \alpha^\prime_{n,t}}_{\text{harvested energy}},
\label{eqn:battery_queue_dynamics}
\end{equation}
with $p_{n,t} \alpha_{n,t} \leq b_{n,t}$.

The system exhibits non-stationary behavior, where a subset of network parameters repeatedly change in a longer time scale compared to the time scale of \gls{SWIPT}. 
In this view, we introduce the notion of a period where the network parameters are considered to be fixed. 
We denote periods by $j$ and each period consists of $T$ time slots. 
Additionally, any given network parameter $x$ at time $t$ in period $j$ is represented by the convention of $x^j_t$. 
In this work, it is assumed that within period $j$, the uncertainties related to the communication and \gls{EH} remain unchanged. 
%More specifically, the power conversion efficiency $\lambda_n$, the scale parameter of the distribution related to \gls{EH} channel $\zeta^\prime_n$, and the scale parameter of the distribution related to communication channels $\zeta_n$, which characterize the \gls{EH} and communication process, are subject to change over certain time intervals.

The objective of the system design is to minimize the time-averaged network energy consumption throughout a given period while ensuring the queue stability by controlling the transmit power and the durations of data transmissions and \gls{EH}. 
For each period $j$, the above goal is formalized as follows
\begin{subequations}\label{eqn:optimization_problem_1}
\begin{align}
\label{eqn:objective_function}
\underset{(\PrimaryTransmissionPower, \EHTime, \PrimaryCommunicationTime, \SecondaryCommunicationTime)_t}{\text{minimize}} \quad &
\lim_{T\to\infty} \frac{1}{T} \sum_{t=0}^T \sum_{n=0}^N \TransmissionPower \CommunicationTime, \\
\label{eqn:time_constraint_1}
\text{subject to} \quad &
0 \leq \EHTime, \PrimaryCommunicationTime, \SecondaryCommunicationTime \leq 1 \quad \forall n \in \mathcal{N}, \forall t, \\
\label{eqn:time_constraint_2} 
& \sum_{n=1}^N\EHTime + \sum_{n=0}^N\SecondaryCommunicationTime \leq 1 \quad \forall t, \\
\label{eqn:queuing_stability_constraint}
& \lim_{T\to\infty} \frac{\sum_{t=0}^T q^j_{n,t}}{T} = 0 \quad \forall n \in \mathcal{N}, \\
\label{eqn:Battery_capacity_constraint}
& 0 \leq b^j_{n,t} \leq B \quad \forall n \in \mathcal{N}, \forall t, \\
\label{eqn:TX1_Power_Constraint}
& 0 \leq \PrimaryTransmissionPower \leq \PrimaryMaxPower \quad \forall t, \\
\label{eqn:TXn_Power_Constraint}
& 0 \leq \alpha_{n,t} p_{n,t} \leq b_{n,t} \quad \forall n\in\mathcal{N}, \\
\label{eqn:queue_dynamics_constraint}
& \eqref{eqn:queue_dynamics}, \eqref{eqn:battery_queue_dynamics}.
\end{align}
\end{subequations}
Constraints~\eqref{eqn:time_constraint_1} and \eqref{eqn:time_constraint_2} ensure feasible scheduling, \eqref{eqn:queuing_stability_constraint} ensures queue stability, while \eqref{eqn:Battery_capacity_constraint} and \eqref{eqn:TX1_Power_Constraint} satisfy energy availability for transmissions. 
Within the context of lifelong machine learning, solving~\eqref{eqn:optimization_problem_1} corresponds to a single \emph{task}, in which, task $j$ is referred to as the process of handling~\eqref{eqn:optimization_problem_1} hereinafter.
Although we define a finite period $j$, during which the network parameters remain fixed, the number of time slots $T$ in a period is assumed to be sufficiently large. 
For example, while individual time slots $t$ may operate on the order of microseconds, a single period $j$ may span several hours. 
In this sense, $T \to \infty$ serves as a mathematical abstraction to facilitate time-averaged analysis within each period.

The overarching goal is to optimize the system’s decisions across all tasks in a non-stationary environment. 
Optimization-based approaches are effective for solving individual tasks under stationary assumptions, they require recomputation whenever network conditions change. 
Therefore, we first present an optimization-based solution to illustrate how a single task (i.e., a fixed period $j$) can be handled using classical methods.
However, since such optimization must be repeated for each period as the environment evolves, this approach becomes computationally inefficient and slow to adapt in practice. 
To overcome this limitation, we next explore a \gls{RL} solution that can learn to adapt through interaction. 
Finally, we introduce a \gls{CD-L2RL} approach, which is specifically designed to continuously adapt to varying network conditions by leveraging knowledge from previously encountered tasks.

\section{Optimization-Based Solution}\label{sec:the Lyapunov_benchmark}

%The optimization problem in~\eqref{eqn:optimization_problem_1} focuses on minimizing the long-term time-averaged energy consumption in a \gls{EH}-\gls{WSN}. 
%In the real-world scenarios of \glspl{WSN} aided by \gls{EH} and as described in the model of this work, the randomness of both the harvested energy and the wireless channels leads to a \textit{non-stationary} environment. 
%Due to this uncertainty, solving optimization problems in such systems with long-term time-averaged objectives and constraints is a significant challenge. 
Although classical \gls{MDP}-based methods can theoretically handle stochastic control problems, they suffer from high computational complexity in large state spaces, making them impractical for real-time operations in large-scale networks.
To address these challenges, we adopt the Lyapunov optimization, a powerful framework for online control in stochastic systems with time-averaged constraints~\cite{qiu.2018}. 
This method provides a low-complexity approach that does not require prior knowledge of system statistics and is well-suited to the energy management needs of dynamic \glspl{WSN}.

To handle time-averaged constraints, we set virtual queues, then define a concatenated vector of the queues as $\Theta_t = [q_{0,t},\ldots,q_{N,t}]$.
Following that, we define the Lyapunov function to measure the congestion in these queues as follows
\begin{equation}\label{the Lyapunov function}
    L(\Theta_t) = \frac{1}{N} \sum_{n=0}^N q_{n,t}^2.
\end{equation}
Then, the Lyapunov drift can be obtained as $\Delta \Theta_t = \mathbb{E} \Biggl\{ L [\Theta_{t+1}] - L [\Theta_{t}] \Biggl| \Theta_t \Biggl\}$. 
According to the Lyapunov optimization theory~\cite{kumar.2022}, enforcing constraints~\eqref{eqn:queue_dynamics_constraint} and~\eqref{eqn:queuing_stability_constraint} is equivalent to minimizing the drift $\Delta \Theta_t$, and optimizing the objective function with the constraints is equivalent to optimizing the drift-plus-penalty defined as $\Delta \Theta_t - \beta \mathbb{E} \bigg[\sum_{n=0}^N \TransmissionPower \CommunicationTime \bigg| \Theta_t\bigg]$, where the parameter $\beta \geq 0$ stands for the penalty weight, which represents the importance of the objective function compared to the constraints at every time slot. 
Therefore, the optimization problem~\eqref{eqn:optimization_problem_1} can be rewritten as follows
\begin{subequations}\label{drift-plus-penalty_opt}
    \begin{equation}
        \begin{aligned}
            \underset{\left(p_0^j, \alpha_n^{\prime j}, \alpha_0^j, \alpha_n^j\right)_t}{\text{min}}~
            & \Delta \Theta_t - \beta \mathbb{E} \bigg[\sum_{n=0}^N p_{n,t}^j \alpha_{n,t}^j \bigg| \Theta_t \bigg] \\
            \text{s.t.}~ 
            & \eqref{eqn:time_constraint_1}, \eqref{eqn:time_constraint_2}, \eqref{eqn:queuing_stability_constraint}, 
            \eqref{eqn:Battery_capacity_constraint}, \eqref{eqn:TX1_Power_Constraint},\eqref{eqn:TXn_Power_Constraint}, 
            \eqref{eqn:queue_dynamics_constraint}, \eqref{eqn:queue_dynamics}, \eqref{eqn:battery_queue_dynamics}.
        \end{aligned}
    \end{equation}
\end{subequations}

According to~\eqref{the Lyapunov function} and the Lyapunov drift, the objective function in~\eqref{drift-plus-penalty_opt} depends on variables at time slot~$t+1$. 
Therefore, the original problem must be reformulated as a sequence of per-slot optimization problems, each solved at time~$t$, to obtain the solution iteratively.
Although the exact drift-plus-penalty expression involves variables at time slot~$t+1$, it can be upper bounded using only the variables at time~$t$. 
Deriving such an upper bound simplifies the optimization problem. 
To this end, we compute an upper bound for the term $q_{n,t+1}^2 - q_{n,t}^2$ as follows
\begin{equation}\label{upper bound Q_1}
    \begin{aligned}
        q_{n,t+1}^2 - q_{n,t}^2 
        &= \left( \max \left\{ q_{n,t} - d_{n,t}, 0 \right\} + a_{n,t} \right)^2 - q_{n,t}^2 \\
        &\leq d_{n,t}^2 + a_{n,t}^2 + 2 q_{n,t}(a_{n,t} - d_{n,t}) \\
        &\leq (1-2 \varrho)d_{n,t}^2 + A^2 + 2 \varrho A,
    \end{aligned}
\end{equation}
where $\varrho$ refers the capacity of data buffers, and $A$ is the maximum arrival data.
From (\ref{upper bound Q_1}), the upper bound of the drift-plus-penalty can be calculated as follows
\begin{equation}\label{upper bound of the drift-plus-penalty}
    \begin{aligned}
    &\Delta \Theta_t - \beta \mathbb{E} \bigg[\sum_{n=0}^N \TransmissionPower \CommunicationTime \bigg| \Theta_t\bigg] \leq \sum_{n=0}^N (1-2 \varrho)d_{n,t}^2 + A^2 + 2 \varrho A - \beta \TransmissionPower \CommunicationTime.
    \end{aligned}
\end{equation}
From~\eqref{upper bound of the drift-plus-penalty}, we observe that the upper bound consists of both variable-dependent terms and constants. 
To design a tractable control policy, we focus only on minimizing the terms that depend on the current decision variables at time slot~$t$, namely $\sum_{n=0}^N (1-2 \varrho)d_{n,t}^2 - \beta \TransmissionPower \CommunicationTime$. 
The remaining terms such as $A^2$ and $2\varrho A$ are constant with respect to the optimization variables and can therefore be omitted from the per-slot problem. 
This leads to the following simplified optimization problem that guides the control decisions at each time slot
\begin{subequations}\label{opt-problem-Final2}
    \begin{equation}
        \begin{aligned}
            \underset{\left(\PrimaryTransmissionPower, \EHTime, \PrimaryCommunicationTime, \SecondaryCommunicationTime\right)_t}{\text{min}}
            &\sum_{n=0}^N (1 - 2 \varrho) d_{n,t}^2 - \beta \TransmissionPower \CommunicationTime \\
            & \text{s.t.}~\eqref{eqn:time_constraint_1}, \eqref{eqn:time_constraint_2}, \eqref{eqn:queuing_stability_constraint}, \eqref{eqn:Battery_capacity_constraint}, \eqref{eqn:TX1_Power_Constraint}, \eqref{eqn:queue_dynamics_constraint}, \eqref{eqn:queue_dynamics}, \eqref{eqn:battery_queue_dynamics}.
        \end{aligned}
    \end{equation}
\end{subequations}
From~\eqref{opt-problem-Final2}, we can affirm that the optimization problem is non-convex due to two key factors. 
First, the data transmission term  $d_{n,t}$ defined in~\eqref{eqn:transmitted_data} is concave, but its square 
$d_{n,t}^2$ appears in the objective, making it a non-convex composition. 
Second, the term  $\TransmissionPower \CommunicationTime$ is bilinear, which is inherently non-convex. 
Therefore, standard convex optimization techniques are not directly applicable.
It is important to emphasize that the original long-term stochastic optimization problem has been decoupled via the Lyapunov optimization framework into a sequence of per-time-slot problems. 
While Eq.~\eqref{opt-problem-Final2} is written in a generic form for all $t$, it serves as a decision-making template that is instantiated and solved separately at each time slot using the current system state. 
%This transformation enables online, adaptive control without requiring knowledge of future system dynamics.
At each time $t$, the observed queue states, battery levels, and channel conditions are substituted into Eq.~\eqref{opt-problem-Final2}, producing a concrete, time-indexed instance of the optimization problem. 
Solving this problem yields the optimal control decisions (e.g., power and time allocations) for that specific slot. 
Because the system state evolves dynamically due to stochastic arrivals and fading, this process must be repeated at every time slot to ensure responsiveness to changing conditions.
While convex relaxation techniques may be employed to approximate a solution~\cite{qiu.2018, wang.2021}, in this paper we adopt an alternating optimization technique~\cite{Gur.2020} to iteratively solve each per-slot problem. 
The full procedure is presented in Algorithm~\ref{alg:optimization_based}, which solves the decoupled per-slot optimization problem at each time $t$ using the observed system state.
\begin{algorithm}[H]
\caption{Optimization-Based Per-Slot Control via Alternating Minimization}
\label{alg:optimization_based}
\begin{algorithmic}
\STATE \textbf{Initialize:} Queue states $q_{n,0}, \forall n \in \{0,\dots, N\}$, battery levels $b_{n,0}, \forall n \in \{1,\dots, N\}$, and other system parameters.
\FOR{each time slot $t = 0, 1, \dots, T$}
    \STATE Observe the current system state $\Theta_t = [q_{0,t}, \dots, q_{N,t}, b_{1,t}, \dots, b_{N,t}]$ and the channel states.
    \STATE Compute the Lyapunov function as defined in~\eqref{the Lyapunov function}

    \STATE Define the drift-plus-penalty objective using the upper bound approximation~\eqref{opt-problem-Final2}
    
    \STATE \textbf{Formulate the per-slot optimization problem} by instantiating Eq.~\eqref{opt-problem-Final2} with the observed system state.
    \STATE \textbf{Solve} the instantiated problem using alternating minimization
    \begin{enumerate}
        \item Initialize control variables with feasible values and set convergence threshold $\epsilon_\mathrm{a}$.
        \item Optimize over $\alpha_{n,t}$ while keeping $(p_{0,t}, \alpha'_{n,t}, \alpha_{0,t})$ fixed.
        \item Optimize over $\alpha_{0,t}$ with $(p_{0,t}, \alpha'_{n,t})$ fixed and the updated $\alpha_{n,t}$.
        \item Optimize over $(p_{0,t}, \alpha'_{n,t})$ with updated values of $\alpha_{0,t}$ and $\alpha_{n,t}$.
        \item Repeat steps (b)–(d) until convergence criterion $\epsilon_\mathrm{a}$ is met.
    \end{enumerate}
    \STATE Apply the resulting control decisions for time slot $t$.
    \STATE Update queue states $q_{n,t+1}, \forall n \in \{0,\dots,N\}$ using Eq.~\eqref{eqn:queue_dynamics}.
    \STATE Update battery states $b_{n,t+1}, \forall n \in \{1,\dots,N\}$ using Eq.~\eqref{eqn:battery_queue_dynamics}.
\ENDFOR
\end{algorithmic}
\end{algorithm}
\section{Reinforcement Learning-Based Solution}\label{sec:RL_benchmark}

Due to the absence of reliable models in dynamic wireless sensor environments, we adopt a model-free \gls{RL} approach that learns optimal policies through direct interaction with the environment~\cite{Aoudia.2018, Ortiz.2017, Atallah.2017, Blasco.2013}. 
This is particularly suitable for our setting, where transition dynamics and reward functions evolve over time. 
To address the need for fine-grained control over continuous-valued actions such as power and time allocations, we employ \gls{PG}-based actor-critic methods~\cite{Peters.2010}, which optimize long-term performance without relying on discretization or explicit models.

\subsection{\gls{MDP} Formulation}

We model the decision-making problem in dynamic \glspl{WSN} as a \gls{MDP}, formally defined as a tuple \(\langle \mathcal{S}, \mathcal{A}, \mathcal{P}, \mathcal{R}, \gamma \rangle\)~\cite{Sutton.1998}, where
%d: 6N
\begin{itemize}
    \item \textbf{State space} $\mathcal{S} \subseteq \mathbb{R}^{6N}$: The system state at the $t$th time step includes, for each $\txI$, the queue length, battery level, communication channel gain, and \gls{EH} channel gain. 
    This results in a $6N$-dimensional state vector as $\boldsymbol{s}_t = [q_{0,t}, \dots, q_{N,t}, b_{1,t}, \dots, b_{N,t}, h_{0,t}, \dots, h_{N,t}, h^\prime_{1,t}, \dots, h'_{N,t}]$.
    
    \item \textbf{Action space} \(\mathcal{A}\): Contains all possible actions where an action \(\boldsymbol{a} = [p_0, \alpha_0, \dots, \alpha_N, \alpha^\prime_1, \dots, \alpha^\prime_N] \in \mathcal{A}\) defines the decision variables, where $\alpha_{n}$ is the fraction of each time slot used for $(TX_n,RX_n)$ data transmission, $\alpha^\prime_{n}$ is the fraction of each time slot used for $(TX_n,RX_n)$ \gls{EH} process, and $p_0$ is the transmission power of $\txPrimary$.

    \item \textbf{Transition function} \(\mathcal{P}(\boldsymbol{s}'|\boldsymbol{s}, \boldsymbol{a})\): Defines the probability distribution over the next state \(\boldsymbol{s}'\), conditioned on the current state \(\boldsymbol{s}\) and action \(\boldsymbol{a}\). Transitions are governed by queue dynamics, battery updates, and stochastic channel variations.

    \item \textbf{Reward function} \(\mathcal{R}(\boldsymbol{s}, \boldsymbol{a})\): The reward function is based on the negative of the objective~\eqref{eqn:optimization_problem_1}, penalizing higher energy consumption, augmented with penalty terms to account for the constraints, satisfying queueing stability. The optimization problem~\eqref{eqn:optimization_problem_1} enforces hard constraints, while the \gls{RL} reward penalizes constraint violations, allowing temporary violations during learning. The reward function is given as $\mathcal{R}(\boldsymbol{s}^j_t, \boldsymbol{a}^j_t) = -\sum_{n=0}^N \TransmissionPower \CommunicationTime - \nu\bigg(\sum_{n=1}^N(\BatteryStateDynamics - B)+\sum_{n=0}^N(\QueueStateDynamics-\TransmittedData)\bigg)$, where \(\nu\) is a shared penalty coefficient applied to both battery and queueing constraints. 
    In a more general dual formulation, each constraint would have its own penalty coefficient (i.e., dual variable), reflecting its individual impact on the objective. 
    In our simulations, we adopt a unified coefficient \(\nu\) as a design simplification that empirically balances these trade-offs effectively.

    \item \textbf{Discount factor} \(\gamma \in [0,1]\): Models the trade-off between immediate and future rewards. We typically use \(\gamma \approx 1\) to prioritize long-term performance.
\end{itemize}
This MDP formulation provides a structured framework for applying \gls{RL} to learn adaptive control strategies under uncertain and time-varying conditions in \glspl{WSN}.
Let $S_d$ and $A_d$ be the dimensionality of the state space and the action space, respectively.

\subsection{Training Algorithm}

As this work involves continuous state and action spaces, we adopt a \gls{PG} method to directly optimize a parameterized control policy via gradient ascent~\cite{Peters.2010}. 
Specifically, we define a stochastic policy $\boldsymbol{\pi}_{\boldsymbol{\theta}}(\boldsymbol{a}^j_t \mid \boldsymbol{s}^j_t)$ that maps states to distributions over actions, where $\boldsymbol{\theta}$ denotes the vector of trainable parameters.
The objective is to maximize the expected cumulative reward over trajectories~\cite{williams.1992}
\begin{equation}\label{eq:RL-objective}
    J(\boldsymbol{\theta}) = \mathbb{E}_{\tau \sim \boldsymbol{\pi}_{\boldsymbol{\theta}}} \left[ \sum_{t=0}^{T} \gamma^t \mathcal{R}(\boldsymbol{s}^j_t, \boldsymbol{a}^j_t) \right],
\end{equation}
where $\tau = (\boldsymbol{s}^j_0, \boldsymbol{a}^j_0, \dots, \boldsymbol{s}^j_T, \boldsymbol{a}^j_T)$ is a \emph{trajectory}, i.e., a sequence of states and actions experienced by the agent when interacting with the environment under policy $\boldsymbol{\pi}_{\boldsymbol{\theta}}$ over a finite horizon $T$.
We model both the policy $\boldsymbol{\pi}_{\boldsymbol{\theta}}(\boldsymbol{a}^j_t \mid \boldsymbol{s}^j_t)$ and the state-value function \(V_{\boldsymbol{\phi}}(\boldsymbol{s}^j_t)\) using \glspl{NN}, parameterized by \(\boldsymbol{\theta}\) and \(\boldsymbol{\phi}\), respectively. 
These networks are trained using \gls{PG} optimization methods, as described below.
Let $N_d$ the number of layers in the neural networks.
To optimize this objective, we apply the \gls{PG} theorem~\cite{Glynn.1990}, which gives
\begin{equation}\label{compute-Gt}
    \nabla_{\boldsymbol{\theta}} J(\boldsymbol{\theta}) = \mathbb{E}_{\tau \sim \boldsymbol{\pi}_{\boldsymbol{\theta}}} \left[ \sum_{t=0}^{T} \nabla_{\boldsymbol{\theta}} \log \boldsymbol{\pi}_{\boldsymbol{\theta}} (\boldsymbol{a}^j_t \mid \boldsymbol{s}^j_t) G_t \right],
\end{equation}
where $G_t = \sum_{t'=t}^{T} \gamma^{t'-t} \mathcal{R}(\boldsymbol{s}^j_{t'}, \boldsymbol{a}^j_{t'})$ is the empirical return from time $t$ onward.
Using raw returns introduces high variance into the gradient estimates. 
To address this, we define the advantage function~\cite{Chung.2021} as $A(\boldsymbol{s}^j_t, \boldsymbol{a}^j_t) = G_t - V(\boldsymbol{s}^j_t)$, where $V(\boldsymbol{s}^j_t)$ is the state-value function learned by a critic network.
Substituting this into the gradient yields a variance-reduced estimator, where $G_t$ will be replaced by $A(\boldsymbol{s}^j_t, \boldsymbol{a}^j_t)$.
This leads to an actor-critic architecture, where the actor updates the policy using gradients and the critic estimates the value function for advantage computation. 
The advantage is computed using the temporal difference (TD) error~\cite{Sutton.1999}
\begin{equation}\label{eq:RL-advantage-Function}
    A(\boldsymbol{s}^j_t, \boldsymbol{a}^j_t) = \mathcal{R}(\boldsymbol{s}^j_t, \boldsymbol{a}^j_t) + \gamma V(\boldsymbol{s}^j_{t+1}) - V(\boldsymbol{s}^j_t).
\end{equation}

As discussed above, we first defined the policy gradient objective, then introduced the baseline to reduce variance, and finally adopted the \gls{TD}-based advantage used in actor-critic training. 
To handle continuous control variables such as transmission power and time allocation, we model the policy using a Gaussian distribution~\cite{Peters.2010} as $\boldsymbol{\pi}_{\boldsymbol{\theta}}(\boldsymbol{a}^j_t \mid \boldsymbol{s}^j_t) = \mathcal{N}(\mu_{\boldsymbol{\theta}}(\boldsymbol{s}^j_t), \Sigma_{\boldsymbol{\theta}}),$ where $\mu_{\boldsymbol{\theta}}(\boldsymbol{s}^j_t)$ and $\Sigma_{\boldsymbol{\theta}}$ are the mean and covariance of the action distribution, parameterized by $\boldsymbol{\theta}$.
The policy parameters are updated using stochastic gradient ascent as $\boldsymbol{\theta} \leftarrow \boldsymbol{\theta} + \alpha \nabla_{\boldsymbol{\theta}} J(\boldsymbol{\theta})$, where $\alpha$ is the learning rate.
We denote the parameters of the value function (critic) as \(\boldsymbol{\phi}\), and the policy (actor) parameters as \(\boldsymbol{\theta}\). 
The critic estimates the state-value function \(V_{\boldsymbol{\phi}}(\boldsymbol{s}^j_t)\) for computing advantages.
The critic is trained by minimizing the squared error between predicted and empirical returns as $\boldsymbol{\phi} \leftarrow \boldsymbol{\phi} - \wp \nabla_{\boldsymbol{\phi}} \left( V_{\boldsymbol{\phi}}(\boldsymbol{s}^j_t) - G_t \right)^2$, where \(\wp\) is the critic learning rate.
The full training procedure is presented in Algorithm~\ref{alg:PG-RL}.
\begin{algorithm}[H]
\caption{\gls{PG}-\gls{RL} Training Algorithm}
\label{alg:PG-RL}
\begin{algorithmic}
\STATE Initialize policy parameters $\boldsymbol{\theta}$ and value function parameters $\boldsymbol{\phi}$
\FOR{each training iteration}
    \STATE Sample a trajectory $\tau = (\boldsymbol{s}^j_0, \boldsymbol{a}^j_0, \dots, \boldsymbol{s}^j_T, \boldsymbol{a}^j_T)$ using policy $\boldsymbol{\pi}_{\boldsymbol{\theta}}$
    \FOR{each time step $t = 0$ to $T$}
        \STATE Compute cumulative return $G_t$ as defined in~\eqref{compute-Gt}
        
        \STATE Estimate advantage using the value function $A(\boldsymbol{s}^j_t, \boldsymbol{a}^j_t)$ as defined in~\eqref{eq:RL-advantage-Function}
        
        \STATE Compute policy gradient $\nabla_{\boldsymbol{\theta}} J(\boldsymbol{\theta})$ as defined in~\eqref{compute-Gt}
        
        \STATE Update policy parameters via gradient ascent $\boldsymbol{\theta} \leftarrow \boldsymbol{\theta} + \alpha \nabla_{\boldsymbol{\theta}} J(\boldsymbol{\theta})$

        \STATE Update value function parameters via gradient descent $\boldsymbol{\phi} \leftarrow \boldsymbol{\phi} - \wp \nabla_{\boldsymbol{\phi}} \left( V_{\boldsymbol{\phi}}(\boldsymbol{s}^j_t) - G_t \right)^2$

    \ENDFOR
\ENDFOR
\end{algorithmic}
\end{algorithm}

\section{Proposed Solution based on A Cross-Domain Lifelong Reinforcement Learning}\label{sec:Proposed_solution}

The overarching objective of this work is to develop a control framework that adaptively optimizes energy and time scheduling policies in \glspl{WSN} deployed in dynamic and heterogeneous environments. 
These environments evolve over time due to variations in external factors such as \gls{EH} efficiency, wireless channel conditions, and data traffic loads. 
This inherent variability gives rise to a sequence of learning problems, each characterized by a specific combination of system parameters and control requirements, for example, a high data arrival rate with low \gls{EH} efficiency, or a low-latency requirement under severe channel fading. 
Such combinations impact how energy and transmission decisions must be made to maintain stable and efficient network operations.
To address this diversity in a structured way, we distinguish between two key concepts: \emph{tasks} and \emph{domains}.

\begin{definition}
A \textbf{\emph{task}} refers to a single learning problem under a fixed environmental configuration. Formally, a task is modeled as a \gls{MDP}, defined by the tuple $\mathcal{F}^{j}=\langle \mathcal{S}, \mathcal{A}, \mathcal{P}^{j}, \mathcal{R}^{j}, \gamma^{j} \rangle$ and $j \in \{1, \ldots, M\}$. 
Each task encapsulates a stationary system behavior over a period and defines a unique objective through its reward function and transition dynamics.
\end{definition}

\begin{definition}
A \textbf{\emph{domain}} is a higher-level abstraction that groups together multiple structurally related tasks. 
Tasks within the same domain share consistent input and output representations (i.e., the same dimensions for $\mathcal{S}$ and $\mathcal{A}$) but may differ in specific parameters such as \gls{EH} conditions, channel profiles, or arrival rates. 
A domain thus captures a family of \glspl{MDP} that operate under similar structural assumptions but distinct environmental regimes. 
For instance, a domain could correspond to a class of \gls{WSN} deployments characterized by shared topologies and action constraints but different \gls{EH} efficiencies or communication noise levels.
\end{definition}

Traditional optimization and \gls{RL} techniques typically focus on single-task or single-domain scenarios, limiting their ability to generalize or adapt to changing conditions. 
To overcome this, we propose a \gls{CD-L2RL} framework, which integrates ideas from both multi-task and cross-domain learning. 
While the proposed solution is grounded in the \gls{MDP} framework, it mitigates high-dimensional complexity by leveraging shared latent structures and sparse policy representations. 
This enables efficient transfer across tasks and domains, avoiding the scalability limitations of classical \gls{MDP} approaches.
The proposed method partitions the overall learning problem into a set of domains, each containing a sequence of tasks. 
Within a domain, task-specific policies are learned as sparse combinations of shared components, enabling efficient intra-domain knowledge transfer. 
Across domains, where state and action spaces may vary, transfer is achieved through domain-specific projection matrices that align heterogeneous tasks with a shared latent representation (i.e., an abstract feature space that captures common structure across tasks).
The \gls{CD-L2RL} framework is designed not only to optimize policies within individual tasks but also to accelerate adaptation in new domains by leveraging previously acquired knowledge. 
The following subsections describe the knowledge transfer mechanism and the architectural components that enable this continual and adaptive learning process.

\subsection{Cross-Domain Knowledge Transfer Mechanism}

To enable \gls{L2RL} across diverse operational regimes, our framework employs a hierarchical knowledge structure that facilitates efficient policy transfer across task domains.
Specifically, we assume a sequence of domains $\mathcal{C}^{1}, \ldots, \mathcal{C}^K$, where each domain $\mathcal{C}^{k}$ is defined by a shared state space $\mathcal{S}^{k}$ and action space $\mathcal{A}^{k}$, along with consistent input-output representations. 
Within each domain, the agent encounters a set of tasks $\mathcal{F}^{j}, \ j \in \{1, \cdots, M\}$, each corresponding to a stationary \gls{MDP} characterized by domain-consistent representations but differing in reward functions or environmental parameters (e.g., \gls{EH} conditions, arrival rates).The goal is to learn a set of optimal task-specific policies
\[
\boldsymbol{\Pi}^\star = \left\{ \boldsymbol{\pi}^{\star}_{\boldsymbol{\theta}_{\text{opt}}^{j}} \right\}_{j=1}^{M}, \quad \text{with parameters} \quad
\boldsymbol{\Theta}^\star = \left\{ \boldsymbol{\theta}_{\text{opt}}^{j} \right\}_{j=1}^{M}.
\]
Within each domain $\mathcal{C}^{k}$, all tasks share a common policy parameter space $\mathbb{R}^{K_d}$, where $K_d = |\mathcal{S}^{k}| \cdot |\mathcal{A}^{k}|$. 
To promote intra-domain knowledge reuse, reusing knowledge between tasks that belong to the same domain, we assume that policy parameters for task $j \in \mathcal{C}^{k}$ are generated via a sparse linear combination over a domain-specific basis matrix $\boldsymbol{G}^{k} \in \mathbb{R}^{K_d \times r}$, where \( r \) denotes the number of latent basis components used to represent task-specific policies within each domain \( \mathcal{C}^k \).
The linear combination over a domain-specific basis matrix is defined as follows
\begin{equation}
\boldsymbol{\theta}^{j} = \boldsymbol{G}^{k} \boldsymbol{v}^{j}, \quad \text{with} \quad \boldsymbol{v}^{j} \in \mathbb{R}^{r},
\end{equation}
where $\boldsymbol{v}^{j}$ is a sparse coefficient vector encoding the contribution of each latent basis to the task's policy. 
Collecting all coefficient vectors in domain $\mathcal{C}^{k}$ gives $\boldsymbol{V}^{(j \in \mathcal{C}^{k})} \in \mathbb{R}^{r \times |\mathcal{C}^{k}|}$.
There is no relation between the domain-specific basis matrix $\boldsymbol{G}^{k}$ and the empirical return $G$. 
There is no relation between coefficient vectors $\boldsymbol{V}^{(j \in \mathcal{C}^{k})}$ and the value functions $V$.

To enable transfer across domains with different input-output structures, we introduce a global shared knowledge repository $\boldsymbol{U} \in \mathbb{R}^{d \times r}$ that contains $r$ latent components applicable across all domains. 
The dimensionality \( d \) defines the size of the shared latent space used for cross-domain knowledge transfer. 
The matrix \( \boldsymbol{U} \in \mathbb{R}^{d \times r} \) serves as a global knowledge base consisting of \( r \) latent components, each of dimension \( d \). 
Each domain $\mathcal{C}^{k}$ is equipped with a projection matrix $\boldsymbol{\Psi}^{\mathcal{C}^{k}} \in \mathbb{R}^{K_d \times d}$ that maps the global latent space to the domain-specific policy space. 
This yields the decomposition $\boldsymbol{G}^{k} = \boldsymbol{\Psi}^{\mathcal{C}^{k}} \boldsymbol{U}$.
Substituting into the parameter formulation gives the cross-domain decomposition as $\boldsymbol{\theta}^{j} = \boldsymbol{\Psi}^{\mathcal{C}^{k}} \boldsymbol{U} \boldsymbol{v}^{j}$.
This decomposition allows each task policy to be synthesized from a small number of shared components, linearly combined via task-specific weights and projected to the appropriate domain representation. 
The sparsity of $\boldsymbol{v}^{j}$ encourages compact task representations and supports transfer by reusing latent structures across both tasks and domains.
The agent sequentially interacts with tasks from various domains without prior knowledge of the number or order of domains or tasks. 
It collects trajectories from each task and incrementally updates the shared knowledge base, domain projections, and sparse codes to refine its generalization capabilities. 
This formulation supports both intra-domain and cross-domain transfer: within a domain, all tasks share the same projection, enabling efficient local adaptation; across domains, the shared latent base $\boldsymbol{U}$ ensures coherence and scalability.
Building upon this foundation, the next subsection details how this knowledge is integrated into the \gls{L2RL} framework,
enabling the agent to adapt efficiently as new tasks arrive sequentially.

\subsection{Cross-Domain Learning Framework}

We propose a \gls{CD-L2RL} framework that enables continual adaptation to new tasks by efficiently reusing shared knowledge across domains. 
%Unlike conventional \gls{RL} methods, which typically assume a single stationary domain, our approach explicitly accounts for domain heterogeneity by disentangling task-specific variability from structural regularities.
\Gls{PG} methods~\cite{Sutton.1999} represent the agent’s policy $\boldsymbol{\pi}$ as a function defined over a vector $\boldsymbol{\theta} \in \mathbb{R}^d$ of control parameters. 
With this parameterized policy, we can compute the optimal parameters $\boldsymbol{\theta}^*$ that maximize the expected average reward as follows
\begin{equation}\label{eqn:expected_return_for_task_j}
    \Gamma(\boldsymbol{\theta}^j) = \mathbb{E} \left[ \frac{1}{T} \sum_{t=1}^T \mathcal{R}^j(\boldsymbol{s}_{t+1}, \boldsymbol{a}_{t+1}) \right] = \int_{\mathbb{H}^j} p_{\boldsymbol{\theta}^j}(\boldsymbol{\tau}) \mathcal{\Re}^j(\boldsymbol{\tau}) d\boldsymbol{\tau}, 
\end{equation}
where $\mathbb{H}^j$ is the set of all possible trajectories, $\mathcal{\Re}^j(\boldsymbol{\tau}) = \frac{1}{T} \sum_{t=1}^T \mathcal{R}^j(\boldsymbol{s}^j_{t+1}, \boldsymbol{a}^j_{t+1})$ is the reward of trajectory $\boldsymbol{\tau}$, and $p_{\boldsymbol{\theta}^j}(\boldsymbol{\tau}) = Q_0(s_0) \prod_{t=0}^{T} p( \boldsymbol{s}^j_{t+1} |  \boldsymbol{s}^j_t, \boldsymbol{a}^j_t) \boldsymbol{\pi}_{\boldsymbol{\theta}^j}(\boldsymbol{a}^j_t | \boldsymbol{s}^j_t)$ is the probability of $\boldsymbol{\tau}$ with initial state distribution $Q_0 : \mathcal{S} \to [0, 1]$. By integrating all possible trajectories $\boldsymbol{\tau}$ weighted by their likelihood $p_{\boldsymbol{\theta}^j}(\boldsymbol{\tau})$ under the policy $\boldsymbol{\pi}_{\boldsymbol{\theta}^j}$, the expected return $\Gamma(\boldsymbol{\theta}^j)$ quantifies the overall performance of the policy for task~$j$.
Given the decomposition $\boldsymbol{\theta}^j = \boldsymbol{\Psi}^{\mathcal{C}^{k}} \boldsymbol{U} \boldsymbol{v}^j$, the objective for lifelong policy learning across domains is given as follows
\begin{equation}
\begin{aligned}\label{eqn:lifelong_learning_loss_function}
L_M&(\boldsymbol{U},\boldsymbol{\Psi}^{\mathcal{C}^{1}},\dots,\boldsymbol{\Psi}^{\mathcal{C}^{k}})= \sum_{k=1}^K\bigg\{ \frac{1}{|\mathcal{C}^{k}|} \sum_{j \in \mathcal{C}^{k}} \min_{\boldsymbol{v}^j} \left[ -\Gamma(\boldsymbol{\theta}^j) + \mu_1 \|\boldsymbol{v}^j\|_1 \right] + \mu_2 \|\boldsymbol{\Psi}^{\mathcal{C}^{k}}\|_F^2 \bigg\} + \mu_3 \|\boldsymbol{U}\|_F^2,
\end{aligned}
\end{equation}
where $\mu_1$, $\mu_2$, and $\mu_3$ are regularization weights.
The $L_1$ norm of $\boldsymbol{v}^{j}$ is used to approximate the true vector sparsity. 
We employ regularization via the Frobenius norm $\|\cdot\|_F$ to avoid overfitting on both the shared knowledge base $\boldsymbol{U}$ and each of the group projections.
To enable efficient policy learning in a lifelong setting, it is necessary to avoid reliance on full trajectory distributions across all tasks. 
The expected return objective for each task $j$, given in Equation~\eqref{eqn:expected_return_for_task_j}, requires access to all possible trajectories, which is impractical for continual learning. 
To address this, we reformulate the lifelong optimization objective using a surrogate that approximates the expected return through a lower-bound-based Taylor expansion.
We begin with the original task-specific objective
\begin{equation}
\min_{\boldsymbol{v}^j} \left[ -\Gamma(\boldsymbol{\theta}^j) + \mu_1 \|\boldsymbol{v}^j\|_1 \right], \quad \text{where} \quad \boldsymbol{\theta}^j = \boldsymbol{\Psi}^{\mathcal{C}^k} \boldsymbol{U} \boldsymbol{v}^j.
\end{equation}
Following the approach in~\cite{Ammar.2015}, we consider the lower bound of $\Gamma(\boldsymbol{\theta}^j)$ by defining
\begin{equation}
\begin{aligned}
\Gamma_{L,\boldsymbol{\theta}}(\tilde{\boldsymbol{\theta}}^j) \propto - \int_{\boldsymbol{\tau} \in \mathbb{H}^j} p_{\boldsymbol{\theta}^j}(\boldsymbol{\tau}) \mathcal{\Re}^j(\boldsymbol{\tau}) \log \left[ \frac{p_{\boldsymbol{\theta}^j}(\boldsymbol{\tau}) \mathcal{\Re}^j(\boldsymbol{\tau})}{p_{\tilde{\boldsymbol{\theta}}^j}(\boldsymbol{\tau})} \right] d\boldsymbol{\tau},
\end{aligned}
\end{equation}
which leads to the equivalent minimization problem
\begin{equation}\label{equevalent_Minimization}
\min_{\tilde{\boldsymbol{\theta}}^j} \int_{\boldsymbol{\tau} \in \mathbb{H}^j} p_{\boldsymbol{\theta}^j}(\boldsymbol{\tau}) \mathcal{\Re}^j(\boldsymbol{\tau}) \log \left[ \frac{p_{\boldsymbol{\theta}^j}(\boldsymbol{\tau}) \mathcal{\Re}^j(\boldsymbol{\tau})}{p_{\tilde{\boldsymbol{\theta}}^j}(\boldsymbol{\tau})} \right] d\boldsymbol{\tau}.
\end{equation}
Since computing this term remains computationally intensive, we adopt an approximation based on a second-order Taylor expansion around a surrogate solution $\boldsymbol{\rho}^j$ defined as $\boldsymbol{\rho}^j = \boldsymbol{\theta}^j + \eta \mathcal{\Im}^{-1} \nabla_{\boldsymbol{\tilde{\theta}}^j} \Gamma_{L,\boldsymbol{\theta}}(\boldsymbol{\tilde{\theta}}^j)$, where $\eta$ is a learning rate and $\mathcal{\Im}$ is the Fisher information matrix.
Using the identity
\begin{equation}
\log p_{\tilde{\boldsymbol{\theta}}^j}(\boldsymbol{\tau}) = \log Q_0(\boldsymbol{s}_0) + \sum_{t=1}^{T^j} \log p(\boldsymbol{s}_{t+1} | \boldsymbol{s}_t, \boldsymbol{a}_t) + \sum_{t=1}^{T^j} \log \boldsymbol{\pi}_{\tilde{\boldsymbol{\theta}}^j}(\boldsymbol{a}_t | \boldsymbol{s}_t),
\end{equation}
the first-order gradient becomes $\nabla_{\tilde{\boldsymbol{\theta}}^j} \Gamma_{L,\boldsymbol{\theta}}(\tilde{\boldsymbol{\theta}}^j) = - \mathbb{E}_{\boldsymbol{\tau} \sim p_{\boldsymbol{\theta}^j}} \left[ \mathcal{\Re}^j(\boldsymbol{\tau}) \sum_{t=1}^{T^j} \nabla_{\tilde{\boldsymbol{\theta}}^j} \log \boldsymbol{\pi}_{\tilde{\boldsymbol{\theta}}^j}(\boldsymbol{a}_t^j | \boldsymbol{s}_t^j) \right]$.
Then, the second derivative is $\boldsymbol{\aleph}^j = - \mathbb{E}_{\boldsymbol{\tau} \sim p_{\boldsymbol{\theta}^j}} \left[ \mathcal{\Re}^j(\boldsymbol{\tau}) \sum_{t=1}^{T^j} \nabla^2_{\tilde{\boldsymbol{\theta}}^j} \log \boldsymbol{\pi}_{\tilde{\boldsymbol{\theta}}^j}(\boldsymbol{a}_t^j | \boldsymbol{s}_t^j) \right]_{\tilde{\boldsymbol{\theta}}^j = \boldsymbol{\rho}^j}$.
We now apply a second-order Taylor expansion of $-\Gamma(\tilde{\boldsymbol{\theta}}^j)$ around $\boldsymbol{\rho}^j$ as
\begin{equation}
\begin{aligned}
-\Gamma(\tilde{\boldsymbol{\theta}}^j) \approx -\Gamma(\boldsymbol{\rho}^j) + (\tilde{\boldsymbol{\theta}}^j - \boldsymbol{\rho}^j)^\top \nabla (-\Gamma)(\boldsymbol{\rho}^j) + \frac{1}{2} (\tilde{\boldsymbol{\theta}}^j - \boldsymbol{\rho}^j)^\top \boldsymbol{\aleph}^j (\tilde{\boldsymbol{\theta}}^j - \boldsymbol{\rho}^j).
\end{aligned}
\end{equation}
Since $\boldsymbol{\rho}^j$ is a stationary point of the lower-bound objective, we have $\nabla (-\Gamma)(\boldsymbol{\rho}^j) = 0$. 
Thus, the linear term vanishes. 
We also discard the constant term $-\Gamma(\boldsymbol{\rho}^j)$ as it does not influence the optimization. 
Substituting $\tilde{\boldsymbol{\theta}}^j = \boldsymbol{\Psi}^{\mathcal{C}^k} \boldsymbol{U} \boldsymbol{v}^j$, we obtain $-\Gamma(\tilde{\boldsymbol{\theta}}^j) \approx \frac{1}{2} \left\| \boldsymbol{\rho}^j - \boldsymbol{\Psi}^{\mathcal{C}^k} \boldsymbol{U} \boldsymbol{v}^j \right\|_{\boldsymbol{\aleph}^j}^2$, where $\| \boldsymbol{x} \|_{\boldsymbol{A}}^2 := \boldsymbol{x}^\top \boldsymbol{A} \boldsymbol{x}$.
Plugging this approximation into the full lifelong objective~\eqref{eqn:lifelong_learning_loss_function} yields $\hat{L}_M(\boldsymbol{U}, \boldsymbol{\Psi}^{\mathcal{C}^{1}}, \dots, \boldsymbol{\Psi}^{\mathcal{C}^{K}})$, where $\Gamma(\boldsymbol{\theta}^j)$ will be replaced by $\Gamma(\tilde{\boldsymbol{\theta}}^j)$.
This formulation eliminates dependence on trajectory sampling while preserving expressiveness and structure in the \gls{CD-L2RL}.

We fit the policy parameters in two stages. 
Upon observing a new task \( j \in \mathcal{C}^{k} \), we first estimate the task-optimal policy parameters \( \boldsymbol{\rho}^j \) via a policy gradient update, along with the associated curvature matrix \( \boldsymbol{\aleph}^j \), which approximates the local Hessian at \( \boldsymbol{\rho}^j \). 
Then, we obtain the task-specific coefficient vector \( \boldsymbol{v}^j \) by solving the following \gls{LASSO}-like optimization to promote sparsity in the solution by penalizing the sum of absolute values of the coefficients
\begin{equation}
\boldsymbol{v}^{j} \leftarrow \arg\min_{\boldsymbol{v}} \left\| \boldsymbol{\rho}^j - \boldsymbol{\Psi}^{\mathcal{C}^k} \boldsymbol{U} \boldsymbol{v} \right\|_{\boldsymbol{\aleph}^{j}}^2 + \mu_1 \|\boldsymbol{v}\|_1.
\end{equation}
To update the shared knowledge components incrementally, we compute the task-specific statistics as $\boldsymbol{X}^{j} = \boldsymbol{v}^{j} (\boldsymbol{v}^{j})^\top, \ \boldsymbol{Y}^{j} = \boldsymbol{\rho}^{j} (\boldsymbol{v}^{j})^\top$.
These statistics are used to recursively update cumulative cross-domain sufficient statistics for each domain as
\begin{equation}
\boldsymbol{X}^{\mathcal{C}^{k}} \leftarrow (1 - \eta_a) \boldsymbol{X}^{\mathcal{C}^{k}} + \eta_a \boldsymbol{X}^{j}, \qquad
\boldsymbol{Y}^{\mathcal{C}^{k}} \leftarrow (1 - \eta_a) \boldsymbol{Y}^{\mathcal{C}^{k}} + \eta_a \boldsymbol{Y}^{j},
\end{equation}
where \( \eta_a \in (0, 1) \) is the update rate controlling the influence of the current task.
Finally, the group-specific projection matrix is updated using the pseudoinverse-based least-squares solution as $\boldsymbol{\Psi}^{\mathcal{C}^{k}} \leftarrow \boldsymbol{Y}^{\mathcal{C}^{k}} \left(\boldsymbol{X}^{\mathcal{C}^{k}}\right)^\dagger$, where \( (\cdot)^\dagger \) denotes the Moore–Penrose pseudoinverse (used to obtain a minimum-norm least-squares solution even when the matrix is not invertible).
This update yields the minimum-norm solution to the least-squares problem $\min_{\boldsymbol{\Psi}^{\mathcal{C}^{k}}} \left\| \boldsymbol{Y}^{\mathcal{C}^{k}} - \boldsymbol{\Psi}^{\mathcal{C}^{k}} \boldsymbol{X}^{\mathcal{C}^{k}} \right\|_F^2$, ensuring stable and consistent refinement of the domain projection matrices even in the presence of noisy or underdetermined data.
Having updated \( \boldsymbol{\Psi}^{\mathcal{C}^{k}} \), the policy parameters for task \( j \) are reconstructed as $\boldsymbol{\theta}^j = \boldsymbol{\Psi}^{\mathcal{C}^{k}} \boldsymbol{U} \boldsymbol{v}^j$, from which the policy \( \boldsymbol{\pi}_{\boldsymbol{\theta}^j} \) can be evaluated and refined through additional local learning if needed.
The proposed \gls{CD-L2RL} algorithm is shown in Algorithm~\ref{alg:CD-L2RL algorithm}.
\begin{algorithm}[H]
\caption{\gls{CD-L2RL} algorithm}
\label{alg:CD-L2RL algorithm}
\begin{algorithmic}
\STATE \textbf{Input:} Regularization parameters $\mu_1$, $\mu_2$, $\mu_3$, learning rate $\eta$
\STATE \textbf{Initialize:} Shared knowledge base, projection matrices, sufficient stats $\{\boldsymbol{X}^{\mathcal{C}^k}, \boldsymbol{Y}^{\mathcal{C}^k}\}$ for each domain
\FOR{each incoming task $j$}
    \STATE Observe task MDP $\mathcal{F}^j = \langle \mathcal{S}, \mathcal{A}, \mathcal{P}^j, \mathcal{R}^j, \gamma^j \rangle$
    \STATE Assign task $j$ to domain $\mathcal{C}^k$
    \STATE Estimate surrogate optimal policy parameters $\boldsymbol{\rho}^j$ and curvature matrix $\boldsymbol{\aleph}^j$ via policy gradient
    \STATE Compute sparse task coefficient $\boldsymbol{v}^{j}$
    
    \STATE Compute sufficient statistics $\boldsymbol{X}^j$ and $\boldsymbol{Y}^j$

    \STATE Update domain-level statistics $\boldsymbol{X}^{\mathcal{C}^k}$ and $\boldsymbol{Y}^{\mathcal{C}^k}$

    \STATE Update projection matrix $\boldsymbol{\Psi}^{\mathcal{C}^k}$ via least-squares fit:

    \STATE Compute policy parameters: $\boldsymbol{\theta}^j = \boldsymbol{\Psi}^{\mathcal{C}^k} \boldsymbol{U} \boldsymbol{v}^j$
\ENDFOR
\STATE \textbf{Output:} Optimal policies $\{ \boldsymbol{\pi}_{\boldsymbol{\theta}^j}^\star \}$ for all tasks $j$
\end{algorithmic}
\end{algorithm}

\section{Simulation Results}\label{sec:simulation_results}

This section evaluates the performance of the the \gls{EH} framework and the proposed algorithms, including \gls{CD-L2RL}, \gls{PG}-\gls{RL}, and the Lyapunov optimization through simulations conducted under dynamic and realistic \gls{WSN} conditions. 
The aim is to benchmark its adaptability and learning performance in comparison with two well-established baselines: the Lyapunov optimization and \gls{PG}-\gls{RL}. 
The experiments are designed to emulate a range of non-stationary scenarios, including time-varying energy harvesting capabilities, stochastic data arrival rates, and fluctuating wireless channel conditions.

The simulation environment consists of multiple sensor nodes communicating wirelessly while harvesting energy. 
To emulate realistic \gls{WSN} behavior, both the information and energy transfer channels are modeled as Rayleigh fading channels. 
The channel gains for data transmission and wireless energy transfer are independently sampled from Rayleigh distributions at each time step.  
In the simulations, $\lambda_n$ is allowed to vary across environments to model real-world fluctuations in \gls{EH} efficiency. 
These aspects directly influence the availability of usable energy and thus affect the control policies learned by the agent.
Data packet arrival rates and channel scales change over time.
The operational parameters used in all scenarios are summarized in Table~\ref{table:simulation_parameters}.
\begin{table}[!t]
\caption{Simulation Parameters}
\centering
\begin{tabular}{|c|c|||c|c|c|}
\hline
\textbf{Parameter} & \textbf{Value} & \textbf{Parameter} & \textbf{Value} \\
\hline
$W$ & 5 MHz & $N_0$ & $-120$ dBm \\
\hline
$P_{1,\text{max}}$ & 300 mW & $B_{\text{max}}$ & 100 mJ \\
\hline
Capacity of data buffers $\varrho$ & $\si{1}{\text{Gbits}}$ & Maximum arrival data $A$ & $\si{100}{\text{Kbits}}$ \\
\hline
$N_d$ & $164$ & $r$ & $5$ \\
\hline
$d$ & $2 r$ & $M$ & $4$ \\
\hline
$\mu_1$, $\mu_2$, $\mu_3$ & $0.1, 0.0001, 0.0001$ & $N$ & $1, 2, 3, 4, 5$ \\
\hline
\end{tabular}
\label{table:simulation_parameters}
\end{table}

\subsection{Non-Stationary System Dynamics}
To evaluate adaptability, the simulations are conducted under changing system conditions that evolve as follows
\begin{itemize}
    \item \textbf{\gls{EH} dynamics:} The number of energy harvesters and their power conversion efficiency change across episodes, simulating sensor deployments in environments with unpredictable \gls{RF} energy availability.
    \item \textbf{Traffic patterns:} Data arrival rates vary between episodes, requiring to adapt scheduling and power allocation strategies to avoid buffer overflows and delays.
    \item \textbf{Channel conditions:} Wireless links fluctuate based on Rayleigh fading profiles.
\end{itemize}
These environmental dynamics define different domains or tasks as illustrated in Table~\ref{tab:env-params}. 
As shown in Table~\ref{tab:env-params}, the performance of the baselines and the proposed solution will be evaluated cross domains and tasks. 
Before analyzing robustness and adaptability under domain shifts, we first benchmark the convergence efficiency and computational complexity of the proposed \gls{CD-L2RL} algorithm against \gls{PG}-\gls{RL} and the Lyapunov optimization.
\begin{table}[!t]
\centering
\caption{Dynamic parameter settings across test environments}
\label{tab:env-params}
\begin{tabular}{|c|c|c|c|c||c|c|c|c|}
\hline
 & \multicolumn{4}{c||}{\textbf{Domain 1}} & \multicolumn{4}{c|}{\textbf{Domain 2}} \\
\hline
 & Task $1$ & Task $2$ & Task $3$ & Task $4$ & Task $1$ & Task $2$ & Task $3$ & Task $4$ \\
\hline
$\zeta^\prime_n$ & 0.1 & 0.2 & 0.3 & 0.4 & 0.2 & 0.5 & 0.8 & 0.9 \\
\hline
$\lambda_n$ & 0.2 & 0.3 & 0.6 & 0.8 & 0.3 & 0.7 & 0.9 & 0.2 \\
\hline
$\zeta_n$ & 0.2 & 0.3 & 0.1 & 0.5 & 0.4 & 0.9 & 0.2 & 0.4 \\
\hline
$\lambda_a$ & 5 & 10 & 15 & 20 & 12 & 25 & 5 & 10 \\
\hline
$N$ & 2 & 2 & 2 & 2 & 4 & 4 & 4 & 4 \\
\hline
\end{tabular}
\end{table}
\subsection{Convergence Behavior}

\begin{figure}
    \centering
    \includegraphics[width=0.6\textwidth]{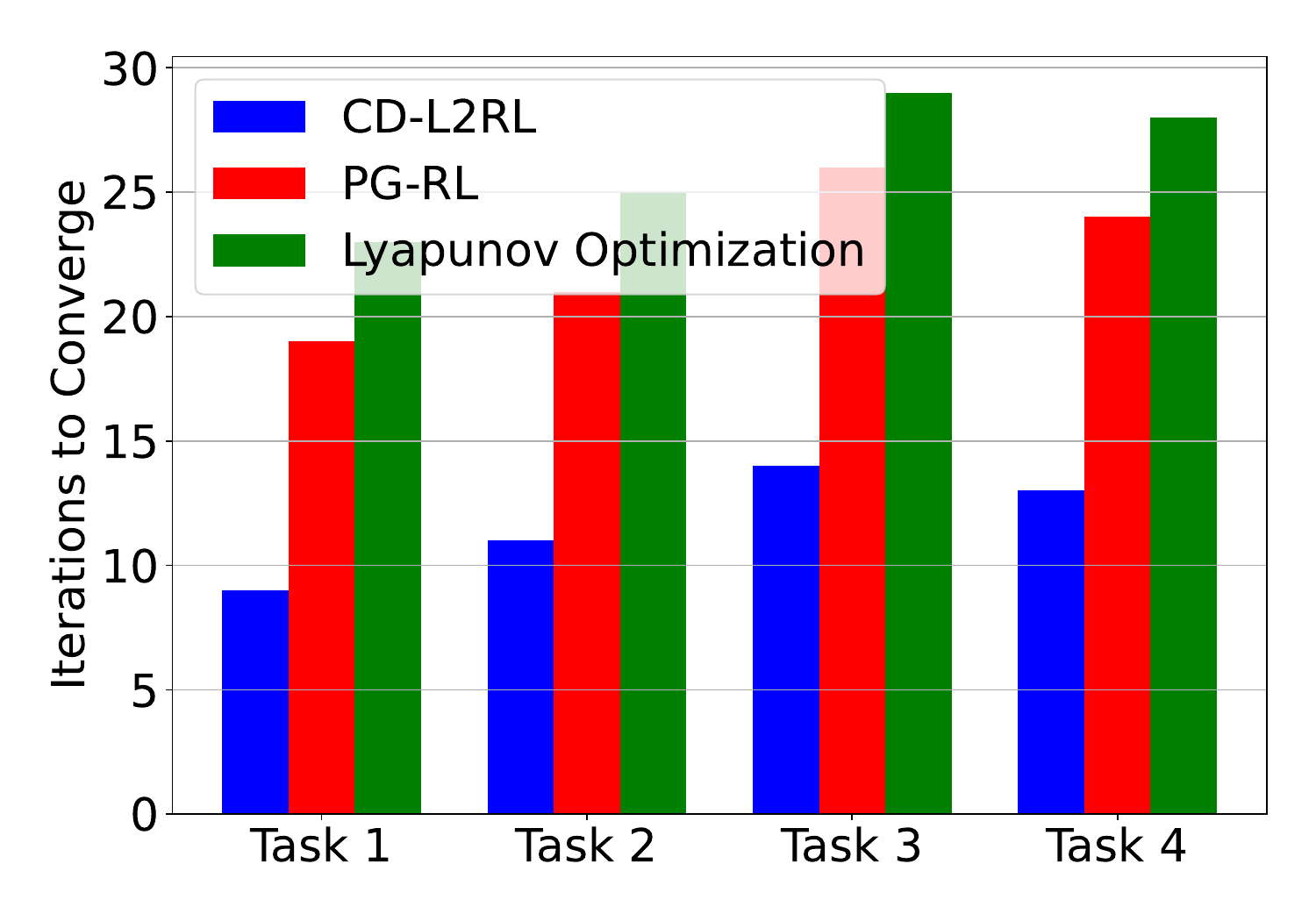}
    \caption{Convergence speed comparison. Bars represent iterations to reach 95\% of maximum reward across four environments.}
    \label{fig:simulation8}
\end{figure}

Fig.~\ref{fig:simulation8} compares the number of iterations required to reach $95\%$ of the maximum average reward across four environments with different dynamic characteristics.
\gls{CD-L2RL} reaches convergence in significantly fewer iterations (between $9$ and $14$) than \gls{PG}-\gls{RL} and the Lyapunov, both of which require more than $20$ iterations. 
The advantage is more pronounced in complex environments, where task structures are dynamic and changing. 
This demonstrates \gls{CD-L2RL}’s ability to generalize knowledge across tasks and environments, significantly reducing training time and enhancing real-time applicability.

\subsection{Complexity Comparison}

This subsection consolidates and compares the computational complexities of the three methods evaluated, the Lyapunov Optimization, \gls{PG}-\gls{RL}, and the proposed \gls{CD-L2RL} algorithm, across algorithmic structure, computational cost, and scalability in dynamic \gls{WSN} environments.

\textbf{The Lyapunov Optimization} operates with low per-slot computational complexity by solving a constrained, non-convex problem through alternating minimization over a small number of control variables (e.g., power and time allocations). 
The per-slot complexity is approximately $\mathcal{O}(N_{\text{alt}} \cdot C_{\text{sub}})$, where $N_{\text{alt}}$ is the number of alternating iterations, and $C_{\text{sub}}$ depends on system dimensionality and the cost of evaluating nonlinear expressions. 
This method scales linearly with the number of nodes but lacks the adaptability required in non-stationary environments.

\textbf{\gls{PG}-\gls{RL}} methods, particularly actor-critic architectures based on policy gradients, offer greater flexibility by learning directly from interaction with the environment. 
They incur high computational cost. 
The complexity per time step is $\mathcal{O}(N_d S_d A_d)$ for both forward and backward passes through the neural networks. 
For $T$ training iterations and accuracy $\epsilon$, the total training complexity becomes $\mathcal{O}(T N_d S_d A_d / \epsilon^2)$. 
Although parallelization and batching improve runtime, the sample inefficiency and need to relearn policies per task make \gls{PG}-\gls{RL} less suitable for lifelong deployment.

\textbf{\gls{CD-L2RL}} achieves a balance between efficiency and adaptability through modular design and hierarchical knowledge sharing. 
The per-task complexity is governed by
\begin{itemize}
    \item $\mathcal{O}(r \cdot K_d)$ for computing task-specific policies from shared representations,
    \item $\mathcal{O}(r \log r)$ for solving the LASSO problem during sparse coding,
    \item $\mathcal{O}(r^2 \cdot d)$ for updating the shared latent base,
    \item $\mathcal{O}(r \cdot K_d^2)$ for updating group-specific projections.
\end{itemize}
Assuming $r, d \ll K_d$, the dominant cost scales linearly in the task dimensionality, making \gls{CD-L2RL} well-suited for real-time, large-scale, and cross-domain learning in dynamic environments.
%A detailed comparison of these approaches is summarized in Table~\ref{tab:comparison}.
Having established the convergence and computational characteristics of the algorithms, we now evaluate their robustness to dynamic \gls{EH} conditions and generalization across tasks and domains.
%
\if0
\begin{table}[!t]
\centering
\caption{Comparison of the Lyapunov Optimization, \gls{PG}-\gls{RL}, and \gls{CD-L2RL}}
\label{tab:comparison}
\resizebox{\linewidth}{!}{
\begin{tabular}{|l|c|c|c|}
\hline
\textbf{Aspect} & \textbf{the Lyapunov Optimization} & \textbf{\gls{PG}-\gls{RL}} & \textbf{\gls{CD-L2RL}} \\
\hline
\textbf{Computational Complexity} & Low & High & Moderate \\
\hline
\textbf{Adaptability} & Low & Medium & High \\
\hline
\textbf{Scalability Across Tasks} & Poor (no reuse) & Poor (relearn each) & Good (knowledge reuse) \\
\hline
\textbf{Best Use Case} & Stationary systems & Adaptive agents & Continual dynamic environments \\
\hline
\end{tabular}
}
\end{table}
%
\fi
\subsection{Discussion of Results}
\begin{figure}
\centering
\includegraphics[width=0.6\textwidth]{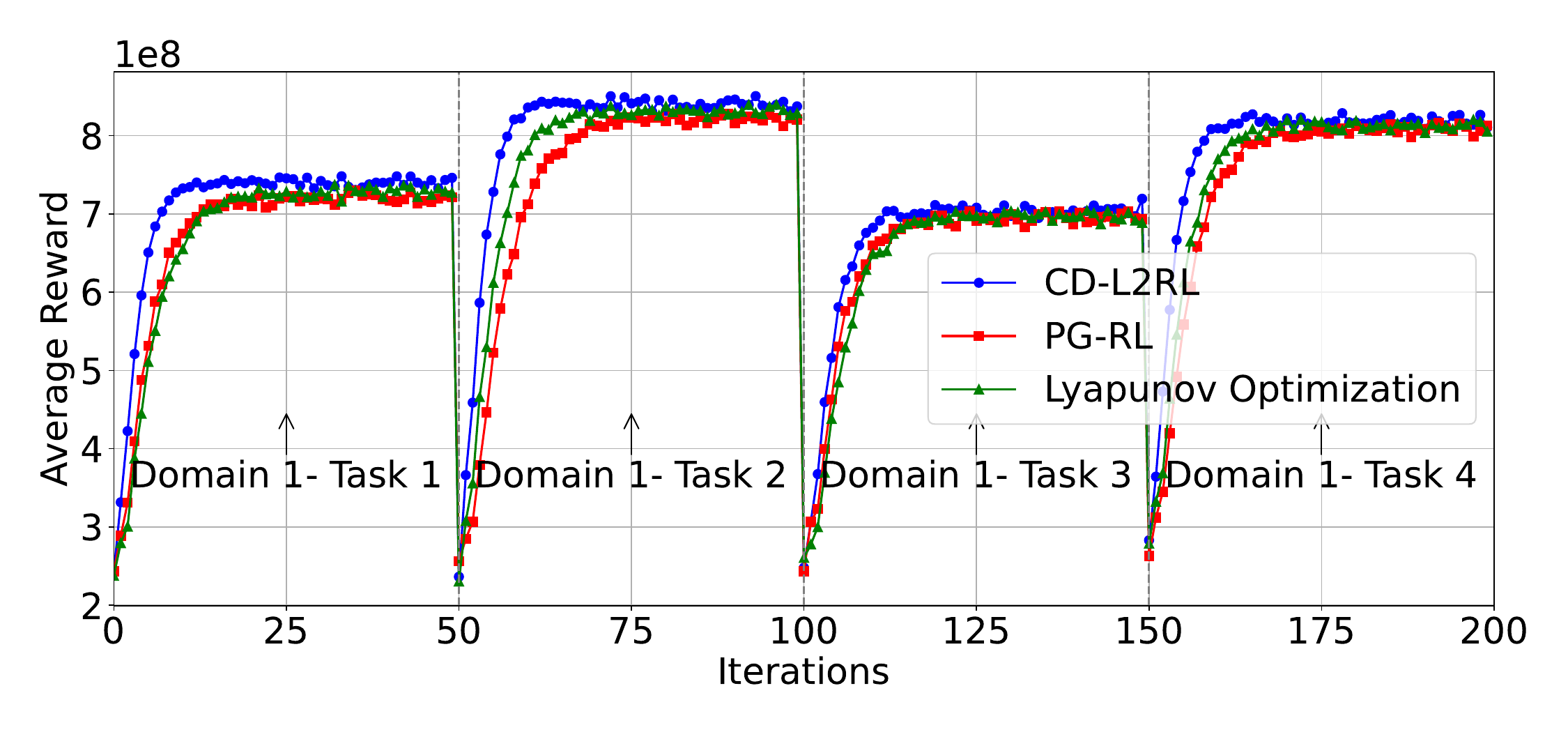}
\caption{The average reward of the Lyapunov optimization, \gls{PG}-\gls{RL}, and our proposed \gls{CD-L2RL} across four different tasks in a single-domain.}
\label{fig:simulation1}
\end{figure}

Fig.~\ref{fig:simulation1} illustrates the average reward achieved by the Lyapunov optimization method, the \gls{PG}-\gls{RL} method, and the proposed \gls{CD-L2RL} method across four tasks in a single-domain. 
Each environment is characterized by a unique configuration of five key dynamic parameters.
As shown in the figure, the proposed \gls{CD-L2RL} method consistently achieves higher average rewards across all tasks, outperforming both baselines significantly. 
This advantage becomes more pronounced in highly dynamic scenarios (e.g., high arrival rate and low conversion efficiency), confirming the method’s ability to generalize and adapt across tasks. 
The reward gap indicates the system’s ability to maintain stable queues under fluctuating input and harvesting conditions, key indicator of robust network operations.

\begin{figure}
    \centering   
    \includegraphics[width=0.6\textwidth]{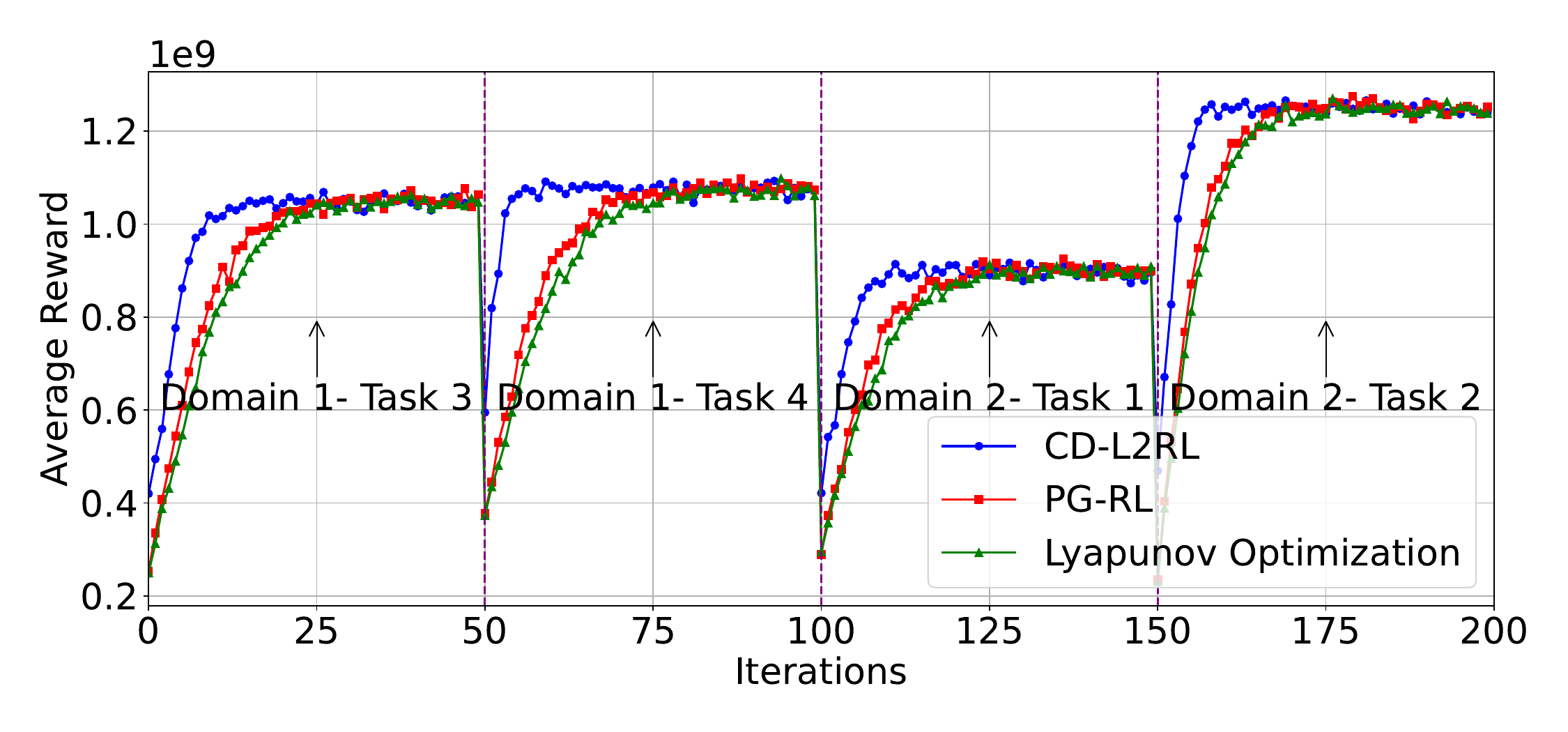}
    \caption{Sequential learning performance of the Lyapunov optimization, \gls{PG}-\gls{RL}, and \gls{CD-L2RL} across domains and tasks.}
    \label{fig:simulation2}
\end{figure}

Fig.~\ref{fig:simulation2} presents the sequential learning trajectories of the Lyapunov optimization, \gls{PG}-\gls{RL}, and the proposed \gls{CD-L2RL} method as the \gls{WSN} transitions through a series of four tasks that are from different domains. 
The vertical lines in the plot denote the boundaries between consecutive environment changes that lead to different tasks and domains.
This setup mimics the real-world operations of \glspl{WSN} that are deployed in continually evolving conditions due to mobility, hardware diversity, or changing application demands. 
The \gls{CD-L2RL} method exhibits remarkable responsiveness to the domain shift. 
It consistently converges to high-performance levels after each transition, requiring significantly fewer samples than its counterparts. 
This is attributed to its use of a shared latent knowledge base, which enables efficient reuse of previously learned policy components. 
In contrast, \gls{PG}-\gls{RL} suffers from slower adaptation, as it must re-learn policies with minimal prior context. 
Additionally, the Lyapunov optimization does not demonstrate effective transfer behavior, resulting in suboptimal transient and steady-state performance.

\begin{figure}
    \centering
    \includegraphics[width=0.6\textwidth]{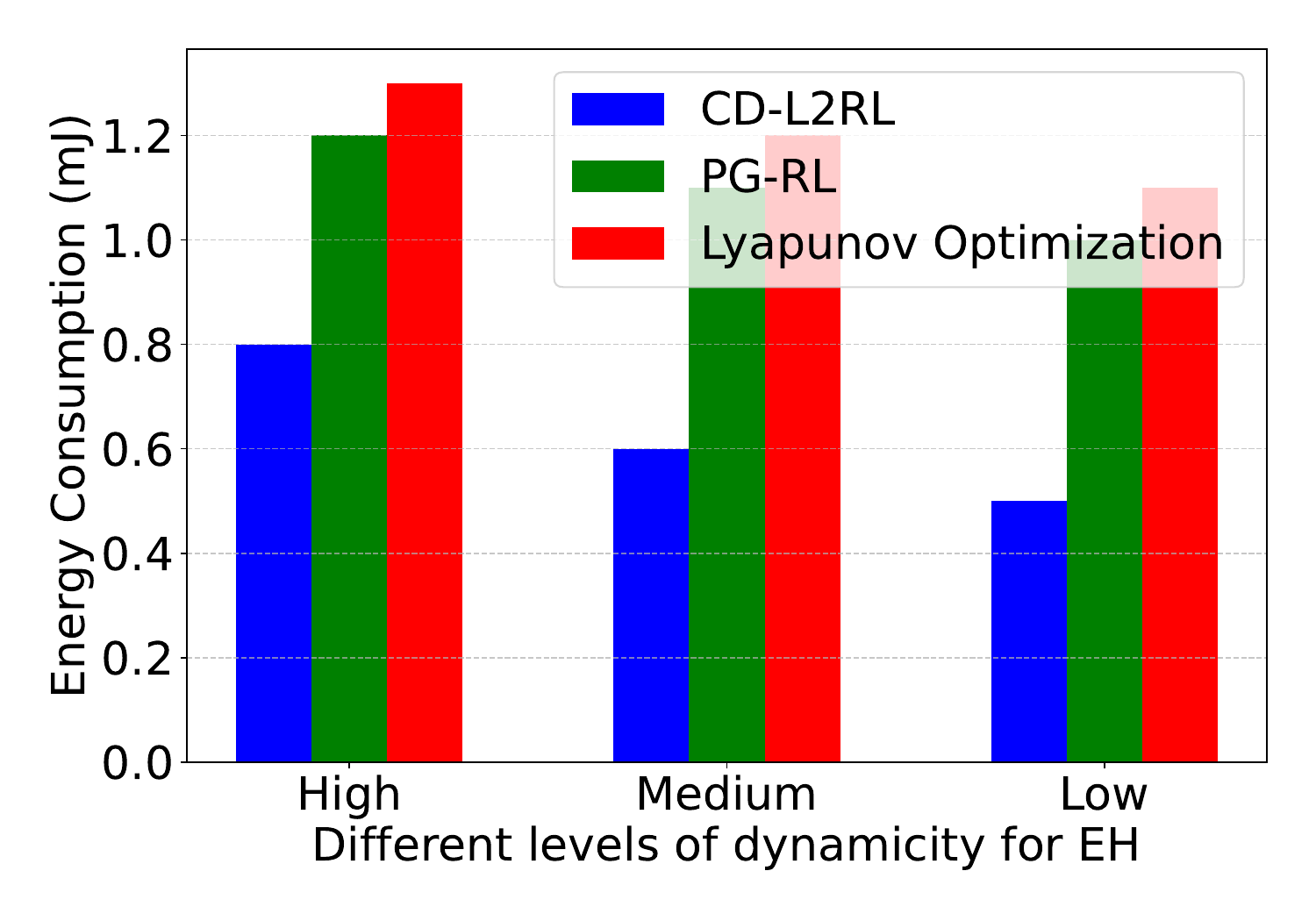}
    \caption{Long-term energy consumption of wireless sensor systems under varying \gls{EH} conditions (quantified by \gls{EH} channel scale and efficiency) for the Lyapunov, \gls{PG}-\gls{RL}, and \gls{CD-L2RL}.}
    \label{fig:simulation3}
\end{figure}

Fig.~\ref{fig:simulation3} evaluates the long-term time-averaged network energy consumption of the Lyapunov optimization, \gls{PG}-\gls{RL}, and \gls{CD-L2RL} methods under varying \gls{EH} dynamics.
We quantify dynamicity based on the \gls{EH} channel scale parameter and power conversion efficiency, which together characterize the variability of harvested energy.
Specifically, the three levels of \gls{EH} dynamicity are defined as follows: \textbf{high dynamicity} corresponds to a channel scale parameter of $1.8$ and an efficiency of $0.65$, \textbf{medium dynamicity} to a channel scale parameter of $1.0$ and an efficiency of $0.45$, and \textbf{low dynamicity} to a channel scale parameter of $0.6$ and an efficiency of $0.35$.
These settings reflect increasingly stable \gls{EH} environments from high to low dynamicity. 
All other system parameters, such as data arrival rate and channel quality, are held constant to isolate the impact of \gls{EH} conditions.
Lower average energy consumption indicates better utilization of available harvested energy and more intelligent power allocation strategies over time.
The proposed \gls{CD-L2RL} algorithm consistently achieves the lowest energy consumption across all \gls{EH} conditions. 
This superior performance results from its ability to learn adaptive power control policies that consider both short-term environmental dynamics and long-term resource sustainability. 
\Gls{CD-L2RL} algorithm effectively aligns power usage with the current harvesting rate and queue state, avoiding overuse during scarce periods and excessive accumulation during abundant periods.
In contrast, the \gls{PG}-\gls{RL} method consumes noticeably more energy due to its slower learning and limited ability to anticipate future energy constraints. 
the Lyapunov optimization, while designed to stabilize queues, relies on static control rules that fail to account for future state dependencies, resulting in inefficient energy usage, especially in fluctuating environments.

\begin{figure}
    \centering
    \includegraphics[width=0.6\textwidth]{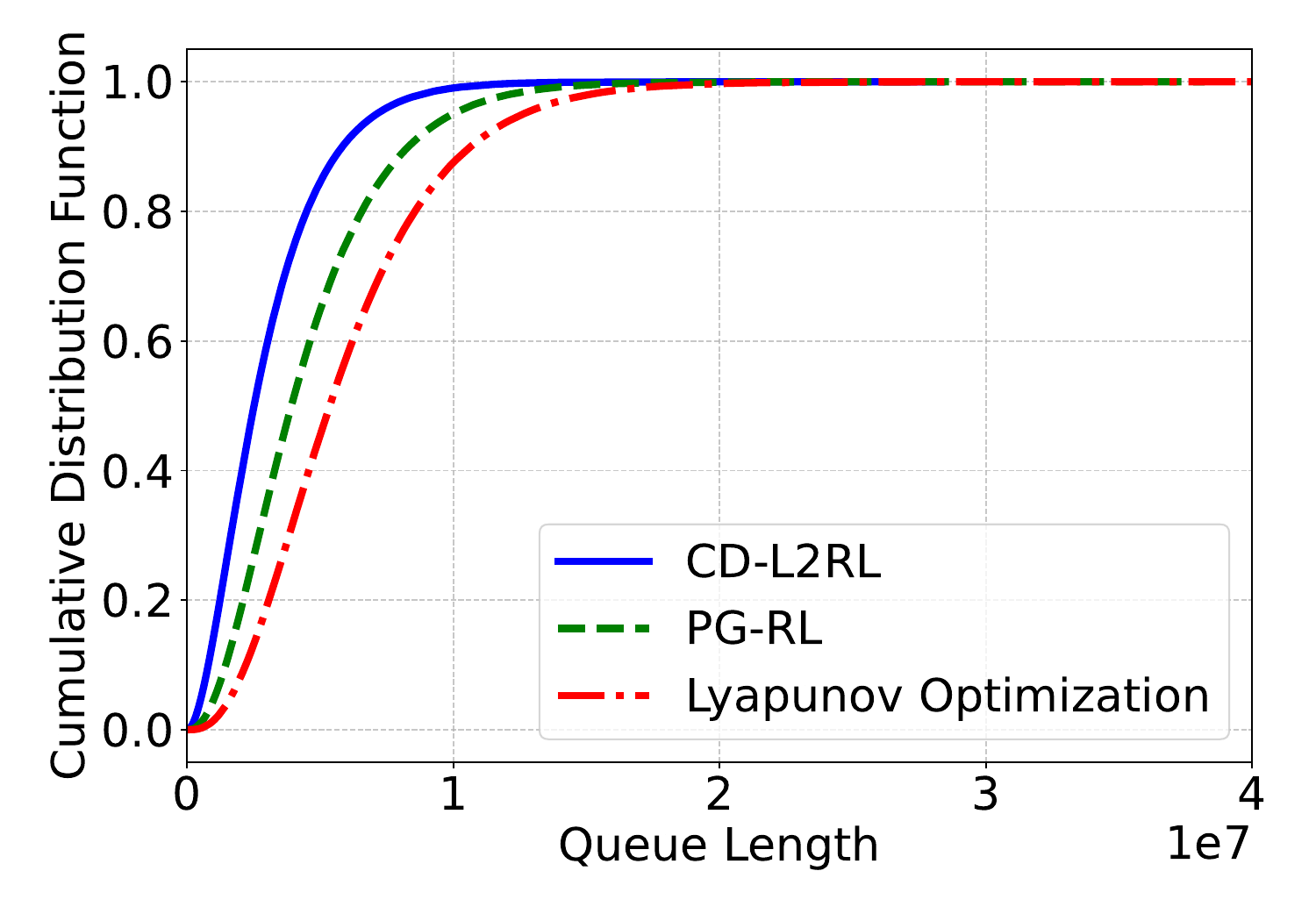}
    \caption{\Gls{CDF} of queue lengths in the second task of the second domain: \gls{CD-L2RL} demonstrates better buffer stability and prevents overflow.}
    \label{fig:simulation4}
\end{figure}

Fig.~\ref{fig:simulation4} illustrates the stability of data queues by presenting the \glspl{CDF} of queue lengths for the Lyapunov, \gls{PG}-\gls{RL}, and \gls{CD-L2RL} methods. 
This analysis is conducted in the second task of the second domain.
The \gls{CDF} plots illustrate the probability that the queue length remains below a specified threshold. 
A steeper \gls{CDF} curve that reaches $1.0$ faster indicates more consistent queue control with lower variance, while flatter curves imply sporadic or large queue backlogs.
As shown in the figure, the \gls{CD-L2RL} algorithm achieves significantly better queue stability than both \gls{PG}-\gls{RL} and the Lyapunov optimization. 
Most of the queue lengths for \gls{CD-L2RL} remain under a tight threshold, indicating the effectiveness of its real-time scheduling and power allocation strategies. 
By anticipating future traffic patterns and adapting to energy availability, \gls{CD-L2RL} maintains buffer occupancy within safe bounds, which is crucial for avoiding packet drops and latency spikes in delay-sensitive \gls{WSN} applications.
In contrast, \gls{PG}-\gls{RL} shows wider spread in queue lengths, reflecting its slower convergence and weaker anticipation of state transitions. 
the Lyapunov optimization, although theoretically queue-aware, lacks a predictive structure and thus exhibits longer tail behavior in the \gls{CDF}, indicating occasional queue build-up and inefficient handling of bursty traffic.

\begin{figure}
    \centering
    \includegraphics[width=0.6\textwidth]{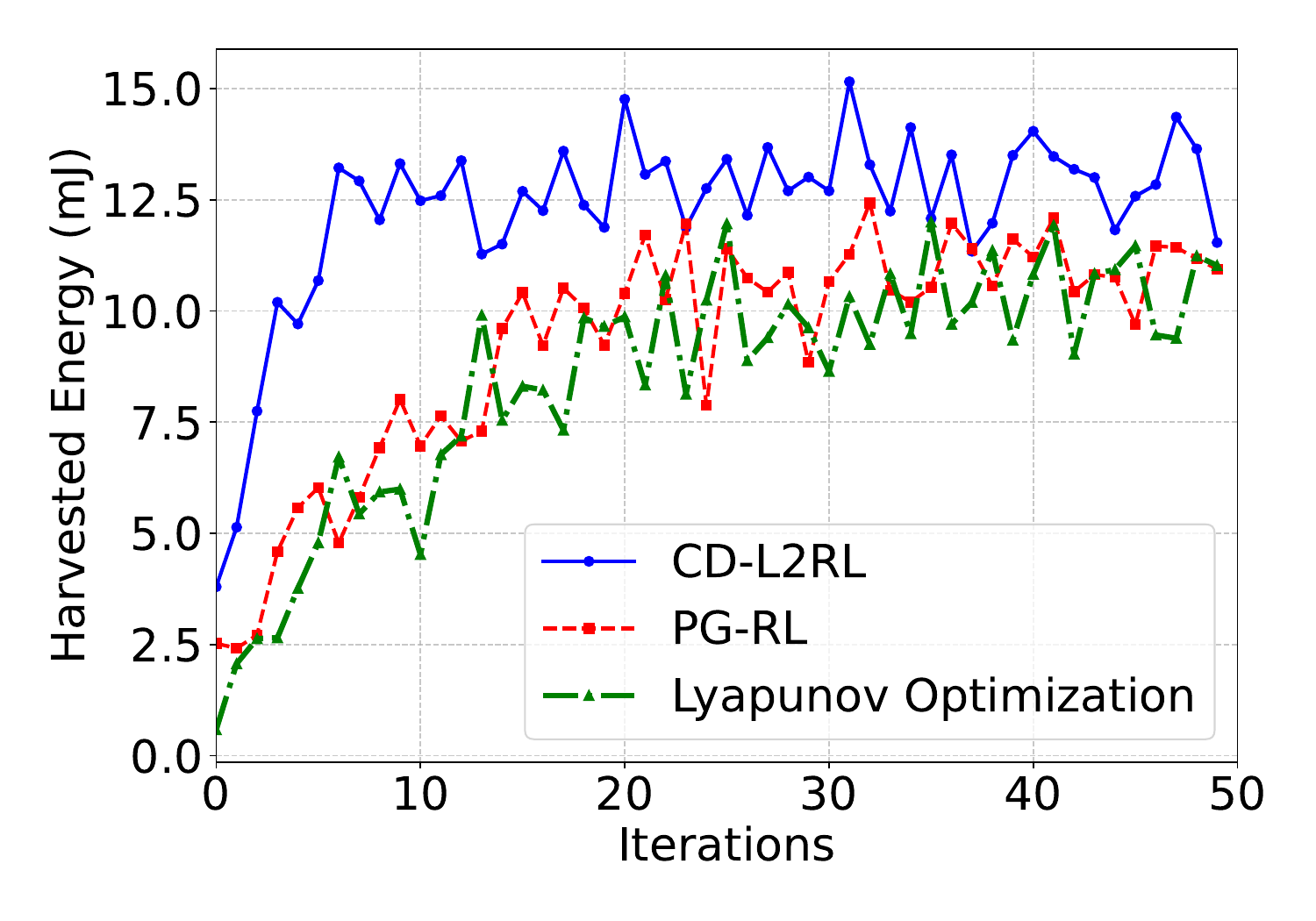}
    \caption{Harvested energy over time in the forth task of the second domain: \gls{CD-L2RL} adapts more effectively to non-stationary energy arrivals.}
    \label{fig:simulation5}
\end{figure}

Fig.~\ref{fig:simulation5} shows the adaptability of the proposed \gls{CD-L2RL} algorithm in highly non-stationary \gls{EH} environments. 
The figure plots the total energy harvested over time by the Lyapunov, \gls{PG}-\gls{RL}, and \gls{CD-L2RL}. 
A desirable algorithm will exhibit a fast and smooth increase in cumulative harvested energy, even as conditions shift.
The \gls{CD-L2RL} method demonstrates superior adaptability, maintaining a steep energy accumulation curve even in the face of all changes. 
This indicates that it not only recovers quickly after a shift but also optimizes harvesting during transitions by reusing and adapting previously learned knowledge. 
Its dynamic policy structure allows it to remain energy-aware across episodes with widely different harvesting profiles.
In contrast, \gls{PG}-\gls{RL} shows intermittent stagnation in harvesting, as it takes time to re-learn effective policies after shift. 
the Lyapunov optimization, which lacks an explicit adaptation mechanism, performs the worst, particularly in low-efficiency environments, as it fails to reallocate harvesting time effectively in response to dynamic changes.
This ability to maintain high cumulative energy yield despite uncertainty is crucial for sustaining operations in long-lived \glspl{WSN} deployed in unpredictable or harsh settings where energy scarcity or volatility is common.

\begin{figure}
    \centering
    \includegraphics[width=0.6\textwidth]{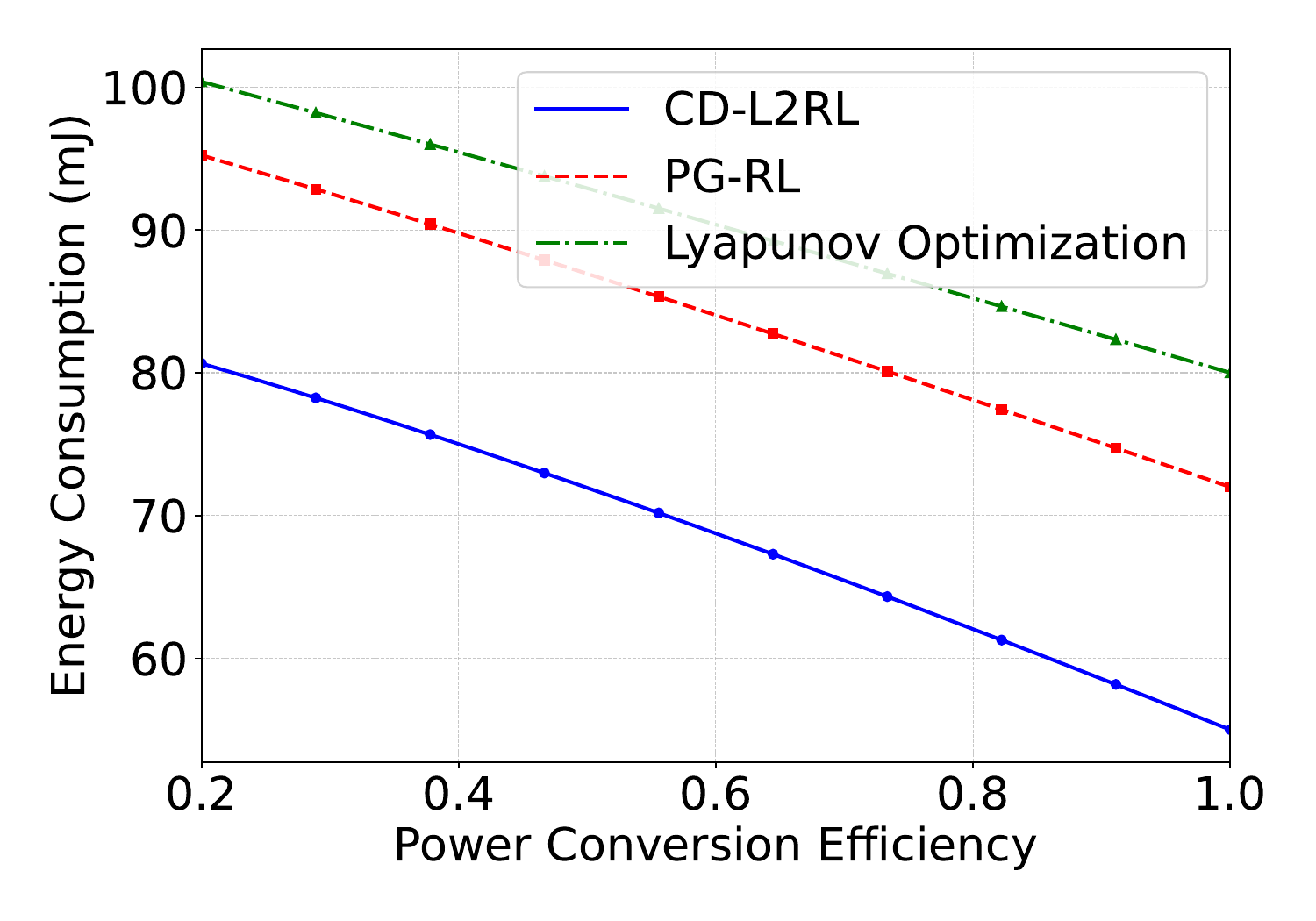}
    \caption{Energy consumption vs. power conversion efficiency. Lower values of $\lambda_n$ imply higher energy loss in conversion.}
    \label{fig:simulation6}
\end{figure}

Fig.~\ref{fig:simulation6} presents the variation in energy consumption as a function of the power conversion efficiency $\lambda_n \in \{0.2, 0.4, 0.6, 0.8, 1.0\}$, which reflects the ratio of received \gls{RF} energy successfully converted into usable electrical energy. 
This parameter models hardware-level imperfections that can significantly impact \gls{EH}. 
The proposed \gls{CD-L2RL} algorithm consistently maintains lower energy consumption across all efficiency values. 
Notably, when $\lambda_n = 0.2$, \gls{CD-L2RL} uses approximately $22\%$ less energy than the Lyapunov optimization method and $18\%$ less than the \gls{PG}-\gls{RL} method. 
As $\lambda_n$ increases, the gap narrows but remains meaningful, demonstrating \gls{CD-L2RL}’s robustness to hardware-level nonidealities. 
This result refers the algorithm's capacity to compensate for inefficient harvesting through intelligent scheduling and control.

\begin{figure}
    \centering
    \includegraphics[width=0.6\textwidth]{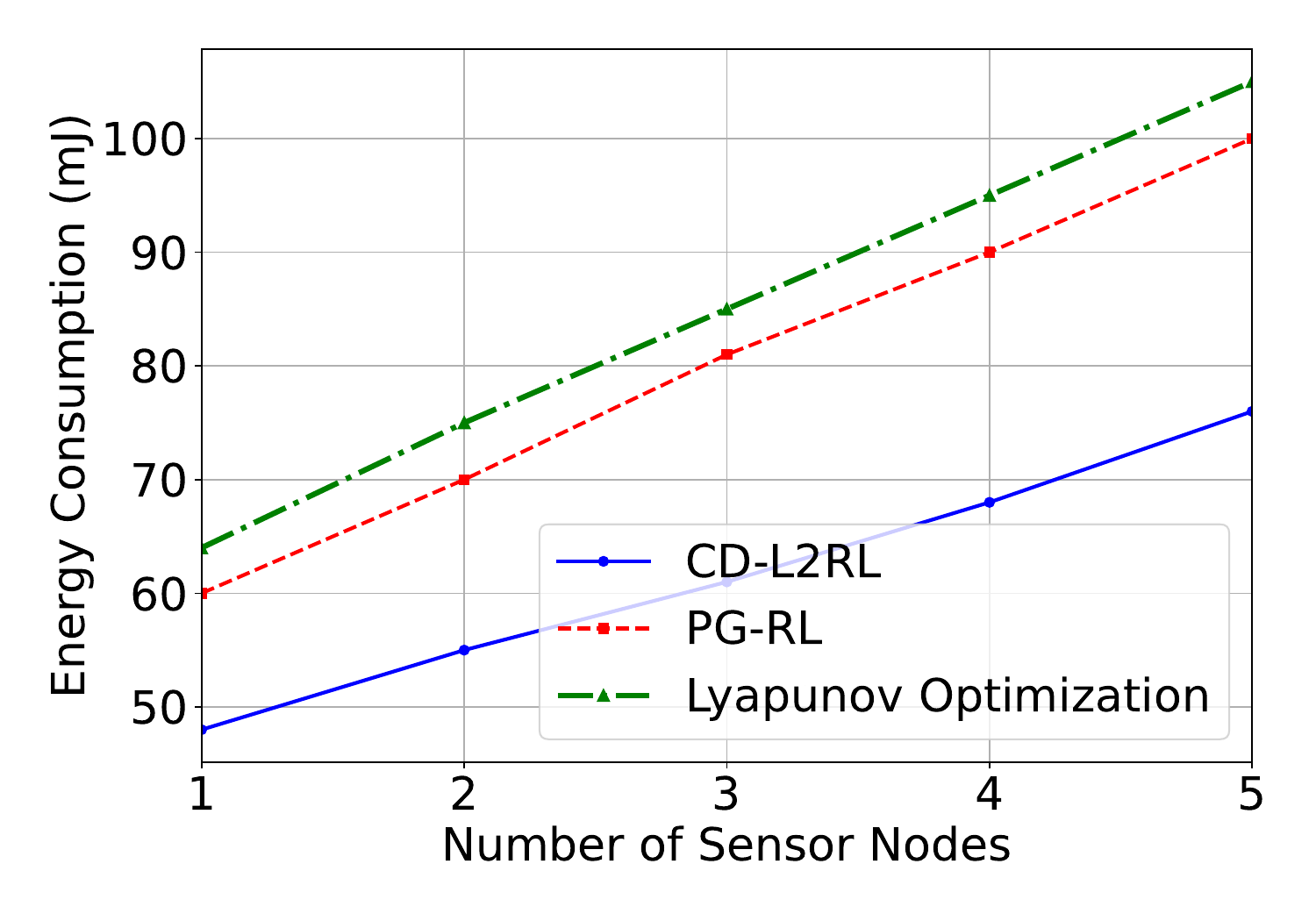}
    \caption{Energy consumption vs. number of sensor nodes. Larger networks test scalability and interference management.}
    \label{fig:simulation7}
\end{figure}

Fig.~\ref{fig:simulation7} examines the impact of scaling the number of sensor nodes $N \in \{1, 2, 3, 4, 5\}$ on energy consumption. 
As $N$ increases, resource contention naturally rises. 
The \gls{CD-L2RL} method exhibits a moderate and near-linear increase in energy usage, maintaining energy efficiency even as the network scales.
In contrast, the energy consumption of \gls{PG}-\gls{RL} and the Lyapunov grows more steeply due to redundant exploration and less adaptive control, respectively. 
This highlights the scalability advantage of \gls{CD-L2RL} in dense deployments. 
Notably, the simulations assume a fixed bandwidth per node, ensuring that the observed improvements are not reliant on proportional bandwidth scaling, but instead reflect algorithmic adaptability under realistic resource constraints.

These figures collectively provide a deep insight into the \gls{CD-L2RL} algorithm’s design and capabilities. 
The results address all critical performance dimensions: convergence, scalability, adaptability, stability, and hardware sensitivity. 
Such a well-rounded evaluation positions \gls{CD-L2RL} as a highly practical solution for intelligent energy management in $6$G \glspl{WSN}.

\section{Conclusion}\label{sec:conclusion_results}
The conclusion of this study underscores the success of the proposed \gls{CD-L2RL} framework in optimizing long-term energy consumption in dynamic \glspl{WSN} environments. 
The \gls{CD-L2RL} algorithm outperforms traditional optimization and \gls{RL} methods by efficiently adapting to non-stationary conditions and leveraging knowledge transfer across both tasks and domains.
By framing dynamic \gls{WSN} control as a multi-task learning problem, \gls{CD-L2RL} learns from a sequence of tasks, each reflecting a unique environmental configuration, while generalizing its experience across multiple domains with structurally related system dynamics.
The simulation results confirm that the \gls{CD-L2RL} approach achieves faster convergence, lower energy consumption, and better adaptability to dynamic environments.
Specifically, the algorithm demonstrates superior performance in maintaining queue stability and minimizing energy consumption under varying \gls{EH} conditions and network traffic scenarios. 
These results highlight the strength of the proposed \gls{CD-L2RL} approach in enabling real-time, sustainable operations of \glspl{WSN} in highly dynamic environments.
The findings suggest that the \gls{CD-L2RL} framework is a promising solution to enhance the sustainability and efficiency of future \glspl{WSN}, paving the way for more resilient and adaptive communication systems in the era of $6$G and beyond. 
Future research will aim to enhance the scalability of the proposed algorithm and explore its integration into distributed systems such as autonomous vehicle networks, smart grids, and edge computing environments.
\bibliographystyle{unsrt} 
\bibliography{Reference}

\begin{thebibliography}{10}

\bibitem{Lu.2013}
Xiao Lu, Ping Wang, Dusit Niyato, Dong~In Kim, and Zhu Han.
\newblock Wireless networks with rf energy harvesting: A contemporary survey.
\newblock {\em IEEE Communications Surveys Tutorials}, 17(2):757--789, 2015.

\bibitem{Yadav.2017}
Animesh Yadav, Mathew Goonewardena, Wessam Ajib, Octavia~A. Dobre, and Halima Elbiaze.
\newblock Energy management for energy harvesting wireless sensors with adaptive retransmission.
\newblock {\em IEEE Transactions on Communications}, 65(12):5487--5498, 2017.

\bibitem{kumar.2022}
Dileep Kumar, Onel L~Alcaraz L{\'o}pez, Satya~Krishna Joshi, and Antti T{\"o}lli.
\newblock Latency-aware multi-antenna swipt system with battery-constrained receivers.
\newblock {\em IEEE Transactions on Wireless Communications}, 22(5):3022--3037, 2022.

\bibitem{Qiu.2019}
Chengrun Qiu, Yang Hu, Yan Chen, and Bing Zeng.
\newblock Deep deterministic policy gradient ({DDPG})-based energy harvesting wireless communications.
\newblock {\em IEEE Internet of Things Journal}, 6(5):8577--8588, 2019.

\bibitem{Ulukus.2015}
Sennur Ulukus, Aylin Yener, Elza Erkip, Osvaldo Simeone, Michele Zorzi, Pulkit Grover, and Kaibin Huang.
\newblock Energy harvesting wireless communications: A review of recent advances.
\newblock {\em IEEE Journal on Selected Areas in Communications}, 33(3):360--381, 2015.

\bibitem{Ozel.2011}
Omur Ozel, Kaya Tutuncuoglu, Jing Yang, Sennur Ulukus, and Aylin Yener.
\newblock Transmission with energy harvesting nodes in fading wireless channels: Optimal policies.
\newblock {\em IEEE Journal on Selected Areas in Communications}, 29(8):1732--1743, 2011.

\bibitem{Ortiz.2017}
Andrea Ortiz, Hussein Al-Shatri, Xiang Li, Tobias Weber, and Anja Klein.
\newblock Reinforcement learning for energy harvesting decode-and-forward two-hop communications.
\newblock {\em IEEE Transactions on Green Communications and Networking}, 1(3):309--319, 2017.

\bibitem{Wang.2015}
Zhe Wang, Vaneet Aggarwal, and Xiaodong Wang.
\newblock Iterative dynamic water-filling for fading multiple-access channels with energy harvesting.
\newblock {\em IEEE Journal on Selected Areas in Communications}, 33(3):382--395, 2015.

\bibitem{Zhang.2019}
Rongrong Zhang, Amiya Nayak, Shurong Zhang, and Jihong Yu.
\newblock Energy-efficient sleep scheduling in wbans: From the perspective of minimum dominating set.
\newblock {\em IEEE Internet of Things Journal}, 6(4):6237--6246, 2019.

\bibitem{Cui.2012}
Ying Cui, Vincent K.~N. Lau, Rui Wang, Huang Huang, and Shunqing Zhang.
\newblock A survey on delay-aware resource control for wireless systems—large deviation theory, stochastic lyapunov drift, and distributed stochastic learning.
\newblock {\em IEEE Transactions on Information Theory}, 58(3):1677--1701, 2012.

\bibitem{Hu.2018}
Yang Hu, Chengrun Qiu, and Yan Chen.
\newblock Lyapunov-optimized two-way relay networks with stochastic energy harvesting.
\newblock {\em IEEE Transactions on Wireless Communications}, 17(9):6280--6292, 2018.

\bibitem{Amirnavaei.2016}
Fatemeh Amirnavaei and Min Dong.
\newblock Online power control optimization for wireless transmission with energy harvesting and storage.
\newblock {\em IEEE Transactions on Wireless Communications}, 15(7):4888--4901, 2016.

\bibitem{Qiu.2018'}
Chengrun Qiu, Yang Hu, and Yan Chen.
\newblock Lyapunov optimized cooperative communications with stochastic energy harvesting relay.
\newblock {\em IEEE Internet of Things Journal}, 5(2):1323--1333, 2018.

\bibitem{qiu.2018}
Chengrun Qiu, Yang Hu, Yan Chen, and Bing Zeng.
\newblock Lyapunov optimization for energy harvesting wireless sensor communications.
\newblock {\em IEEE Internet of Things Journal}, 5(3):1947--1956, 2018.

\bibitem{Li.2015}
Wei Li, Meng-Lin Ku, Yan Chen, and K.~J.~Ray Liu.
\newblock On outage probability for stochastic energy harvesting communications in fading channels.
\newblock {\em IEEE Signal Processing Letters}, 22(11):1893--1897, 2015.

\bibitem{Li.2016}
Wei Li, Meng-Lin Ku, Yan Chen, and K.~J. Ray~Liu.
\newblock On outage probability for two-way relay networks with stochastic energy harvesting.
\newblock {\em IEEE Transactions on Communications}, 64(5):1901--1915, 2016.

\bibitem{Gong.2018}
Jie Gong, Zhenyu Zhou, and Sheng Zhou.
\newblock On the time scales of energy arrival and channel fading in energy harvesting communications.
\newblock {\em IEEE Transactions on Green Communications and Networking}, 2(2):482--492, 2018.

\bibitem{Li.2017}
Wei Li, Meng-Lin Ku, Yan Chen, K.~J.~Ray Liu, and Shihua Zhu.
\newblock Performance analysis for two-way network-coded dual-relay networks with stochastic energy harvesting.
\newblock {\em IEEE Transactions on Wireless Communications}, 16(9):5747--5761, 2017.

\bibitem{Ku.2015'}
Meng-Lin Ku, Wei Li, Yan Chen, and K.~J. Ray~Liu.
\newblock On energy harvesting gain and diversity analysis in cooperative communications.
\newblock {\em IEEE Journal on Selected Areas in Communications}, 33(12):2641--2657, 2015.

\bibitem{Ku.2015}
Meng-Lin Ku, Yan Chen, and K.~J.~Ray Liu.
\newblock Data-driven stochastic models and policies for energy harvesting sensor communications.
\newblock {\em IEEE Journal on Selected Areas in Communications}, 33(8):1505--1520, 2015.

\bibitem{Aoudia.2018}
Fayçal Ait~Aoudia, Matthieu Gautier, and Olivier Berder.
\newblock ({RLMan}): An energy manager based on reinforcement learning for energy harvesting wireless sensor networks.
\newblock {\em IEEE Transactions on Green Communications and Networking}, 2(2):408--417, 2018.

\bibitem{Atallah.2017}
Ribal~F. Atallah, Chadi~M. Assi, and Jia~Yuan Yu.
\newblock A reinforcement learning technique for optimizing downlink scheduling in an energy-limited vehicular network.
\newblock {\em IEEE Transactions on Vehicular Technology}, 66(6):4592--4601, 2017.

\bibitem{Blasco.2013}
Pol Blasco, Deniz Gunduz, and Mischa Dohler.
\newblock A learning theoretic approach to energy harvesting communication system optimization.
\newblock {\em IEEE Transactions on Wireless Communications}, 12(4):1872--1882, 2013.

\bibitem{Cao.2019}
Bin Cao, Long Zhang, Yun Li, Daquan Feng, and Wei Cao.
\newblock Intelligent offloading in multi-access edge computing: A state-of-the-art review and framework.
\newblock {\em IEEE Communications Magazine}, 57(3):56--62, 2019.

\bibitem{Wei.2018}
Yifei Wei, F.~Richard Yu, Mei Song, and Zhu Han.
\newblock User scheduling and resource allocation in hetnets with hybrid energy supply: An actor-critic reinforcement learning approach.
\newblock {\em IEEE Transactions on Wireless Communications}, 17(1):680--692, 2018.

\bibitem{Timothy.2016}
Timothy~P. Lillicrap, Jonathan~J. Hunt, Alexander Pritzel, Nicolas Heess, Tom Erez, Yuval Tassa, David Silver, and Daan Wierstra.
\newblock Continuous control with deep reinforcement learning.
\newblock In Yoshua Bengio and Yann LeCun, editors, {\em 4th International Conference on Learning Representations, {ICLR} 2016, San Juan, Puerto Rico, May 2-4, 2016, Conference Track Proceedings}, 2016.

\bibitem{Firouzjaei.2025}
Hossein~Mohammadi Firouzjaei, Rafaela Scaciota, and Sumudu Samarakoon.
\newblock Multi-task lifelong reinforcement learning for wireless sensor networks.
\newblock {\em arXiv preprint arXiv:2506.16254}, 2025.

\bibitem{Ammar.2014}
Haitham~Bou Ammar, Eric Eaton, Paul Ruvolo, and Matthew~E. Taylor.
\newblock Online multi-task learning for policy gradient methods.
\newblock In {\em Proceedings of the 31st International Conference on International Conference on Machine Learning - Volume 32}, ICML'14, page II–1206–II–1214. JMLR.org, 2014.

\bibitem{kumar.2012}
Abhishek Kumar and Hal Daume~III.
\newblock Learning task grouping and overlap in multi-task learning.
\newblock {\em arXiv preprint arXiv:1206.6417}, 2012.

\bibitem{Ammar.2015}
Haitham~Bou Ammar, Eric Eaton, Jos\'{e}~Marcio Luna, and Paul Ruvolo.
\newblock Autonomous cross-domain knowledge transfer in lifelong policy gradient reinforcement learning.
\newblock In {\em Proceedings of the 24th International Conference on Artificial Intelligence}, IJCAI'15, page 3345–3351. AAAI Press, 2015.

\bibitem{Han.2012}
Shaobo Han, Xuejun Liao, and Lawrence Carin.
\newblock Cross-domain multitask learning with latent probit models.
\newblock {\em arXiv preprint arXiv:1206.6419}, 2012.

\bibitem{wang.2021}
Tong Wang, Yang Shen, Lin Gao, Yufei Jiang, Xu~Zhu, and Fu-Chun Zheng.
\newblock Long-term energy consumption and transmission delay tradeoff in wireless-powered body area networks.
\newblock {\em IEEE Internet of Things Journal}, 9(6):4051--4064, 2021.

\bibitem{Gur.2020}
Eyal Gur, Shoham Sabach, and Shimrit Shtern.
\newblock Alternating minimization based first-order method for the wireless sensor network localization problem.
\newblock {\em IEEE Transactions on Signal Processing}, 68:6418--6431, 2020.

\bibitem{Peters.2010}
Jan Peters.
\newblock Policy gradient methods.
\newblock {\em Scholarpedia}, 5:3698, 01 2010.

\bibitem{Sutton.1998}
R.S. Sutton and A.G. Barto.
\newblock Reinforcement learning: An introduction.
\newblock {\em IEEE Transactions on Neural Networks}, 9(5):1054--1054, 1998.

\bibitem{williams.1992}
Ronald~J. Williams.
\newblock Simple statistical gradient-following algorithms for connectionist reinforcement learning.
\newblock {\em Machine Learning}, 8(3):229--256, May 1992.

\bibitem{Glynn.1990}
Peter~W. Glynn.
\newblock Likelihood ratio gradient estimation for stochastic systems.
\newblock {\em Commun. ACM}, 33(10):75–84, October 1990.

\bibitem{Chung.2021}
Wesley Chung, Valentin Thomas, Marlos~C. Machado, and Nicolas~Le Roux.
\newblock Beyond variance reduction: Understanding the true impact of baselines on policy optimization.
\newblock In Marina Meila and Tong Zhang, editors, {\em Proceedings of the 38th International Conference on Machine Learning}, volume 139 of {\em Proceedings of Machine Learning Research}, pages 1999--2009. PMLR, 18--24 Jul 2021.

\bibitem{Sutton.1999}
Richard~S Sutton, David McAllester, Satinder Singh, and Yishay Mansour.
\newblock Policy gradient methods for reinforcement learning with function approximation.
\newblock In S.~Solla, T.~Leen, and K.~M\"{u}ller, editors, {\em Advances in Neural Information Processing Systems}, volume~12. MIT Press, 1999.

\end{thebibliography}

\end{document}